\documentclass[manuscript]{aastex63}

\newcommand\aastex{AAS\TeX}

\usepackage{natbib}
\usepackage{amsmath}
\usepackage{amssymb}
\usepackage{amsfonts}
\usepackage{amsthm}

\received{December 9, 2019}
\revised{April 2, 2020 }
\accepted{April 18, 2020}
\submitjournal{ApJ}

\shorttitle{\aastex\ A solar magnetic-fan flaring arch}
\shortauthors{Lee et al.}

\begin{document}

\title{A Solar Magnetic-fan Flaring Arch Heated by Non-thermal Particles and Hot Plasma from an X-ray Jet Eruption}

\correspondingauthor{Kyoung-Sun Lee}
\email{kyoungsun.lee@uah.edu, kyoungsun.lee@nasa.gov}

\author[0000-0002-4329-9546]{Kyoung-Sun Lee}
\affil{Center for Space Plasma and Aeronomic Research, University of Alabama in Huntsville, 320 Sparkman Drive, Huntsville, AL 35805, USA}

\author[0000-0001-5686-3081]{Hirohisa Hara}
\affil{Solar Science Observatory, National Astronomical Observatory of Japan, 2-21-1 Osawa, Mitaka, Tokyo, 181-8588, Japan}

\author[0000-0003-0321-7881]{Kyoko Watanabe}
\affil{National Defense Academy of Japan, 1-10-20 Hashirimizu, Yokosuka 239-8686, Japan}

\author{Anand D. Joshi}
\affil{Solar Science Observatory, National Astronomical Observatory of Japan, 2-21-1 Osawa, Mitaka, Tokyo, 181-8588, Japan}

\author[0000-0002-2189-9313]{David H. Brooks}
\affil{College of Science, George Mason University, 4400 University Drive, Fairfax, VA 22030, USA}

\author[0000-0001-7891-3916]{Shinsuke Imada}
\affil{Institute for Space-Earth Environmental Research, Nagoya University, Furo-cho, Chikusa-ku, Nagoya 466-8550, Japan}

\author[0000-0003-0819-464X]{Avijeet Prasad}
\affil{Center for Space Plasma and Aeronomic Research, University of Alabama in Huntsville, 320 Sparkman Drive, Huntsville, AL 35805, USA}

\author{Phillip, Dang}
\affil{Data Scientist at JDA, 500 E John Carpenter Fwy, Irving, TX 75062, USA}

\author[0000-0003-4764-6856]{Toshifumi Shimizu}
\affil{Hinode Team, ISAS/JAXA, 3-1-1 Yoshinodai, Chuo-ku, Sagamihara, Kanagawa 252-5210, Japan}

\author[0000-0002-6172-0517]{Sabrina L. Savage}
\affil{NASA Marshall Space Flight Center, ST 13, Huntsville, AL 35812, USA}

\author[0000-0002-5691-6152]{Ronald Moore}
\affil{Center for Space Plasma and Aeronomic Research, University of Alabama in Huntsville, 320 Sparkman Drive, Huntsville, AL 35805, USA}
\affil{NASA Marshall Space Flight Center, ST 13, Huntsville, AL 35812, USA}

\author[0000-0001-7620-362X]{Navdeep K. Panesar}
\affil{Lockheed Martin Solar and Astrophysics Laboratory, 3251 Hanover Street, Building 252, Palo Alto, CA 94304, USA}
\affil{Bay Area Environmental Research Institute, NASA Research Park, Moffett Field, CA 94035, USA}

\author[0000-0003-4739-1152]{Jeffrey W. Reep}
\affil{Space Science Division, Naval Research Laboratory, Washington, DC 20375, USA}

\begin{abstract}

We have investigated an M1.3 limb flare, which develops as a magnetic loop/arch that fans out from an X-ray jet. Using {\it Hinode}/EIS, we found that the temperature increases with height to a value of over 10$^{7}$\,K at the loop-top during the flare. The measured Doppler velocity (redshifts of 100$-$500\,km\,s$^{-1}$) and the non-thermal velocity ($\geq$100\,km~s$^{-1}$) from \ion{Fe}{24} also increase with loop height. The electron density increases from $0.3\times10^{9}$\,cm$^{-3}$ early in the flare rise to $1.3\times10^{9}$\,cm$^{-3}$ after the flare peak. The 3-D structure of the loop derived with {\it STEREO}/EUVI indicates that the strong redshift in the loop-top region is due to upflowing plasma originating from the jet. Both hard X-ray and soft X-ray emission from {\it RHESSI} were only seen as footpoint brightenings during the impulsive phase of the flare, then, soft X-ray emission moves to the loop-top in the decay phase. Based on the temperature and density measurements and theoretical cooling models, the temperature evolution of the flare arch is consistent with impulsive heating during the jet eruption followed by conductive cooling via evaporation and minor prolonged heating in the top of the fan loop. Investigating the magnetic field topology and squashing factor map from {\it SDO}/HMI, we conclude that the observed magnetic-fan flaring arch is mostly heated from low atmospheric reconnection accompanying the jet ejection, instead of from reconnection above the arch as expected in the standard flare model.
 
\end{abstract}

\keywords{Sun: activity --- Sun: corona --- Sun: flares --- Sun: UV radiation --- techniques: imaging spectroscopy}

\section{Introduction} \label{sec:intro}

Solar flares release explosive energy that results in immense instantaneous heating of the plasma in the solar atmosphere. The standard solar flare model (CSHKP; \citet{carmichael_1964, sturrock_1968, hirayama_1974, kopp&pneuman_1976}) proposes that magnetic reconnection occurs at coronal heights, and the subsequently released magnetic energy is transported to the lower atmospheric layers by several mechanisms, such as thermal conduction or non-thermal particles. Previous observations from soft X-ray (SXR) and hard X-ray (HXR) imagers and spectrographs, such as the {\it Yohkoh}/Soft X-ray Telescope (SXT, \citet{tsuneta_etal1991}) and the {\it Reuven Ramaty High Energy Solar Spectroscopic Imager} ({\it RHESSI}, \citet{lin_etal2002}), reported cusp-shaped, hot loop-top regions ($\sim$10 MK). The associated strong non-thermal emission provides indirect evidence of magnetic reconnection and particle acceleration occuring near the loop-top region \citep{masuda_etal1994, sui&holman_2003, krucker_etal2008, narukage_etal2014}. \citet{sui&holman_2003}, \citet{sui_etal2004}, and \citet{krucker_etal2008} demonstrated the height variation of the HXR emitting region during flares and suggested that the HXR emission from the loop-top region implied reconnection in the corona. \citet{aschwanden_etal1996} studied HXR emissions at different energies using the time-of-flight (TOF) method, which calculates the timing relationship with loop distance and estimates the electron acceleration location above the flaring loops. \citet{narukage_etal2014} presented a direct imaging observation of a hot, non-thermal loop-top source using {\it{Yohkoh}} and the Nobeyama Radioheliograph (NoRH, \citet{nakajima_etal1994}), again suggesting that the reconnection point could be above the flare loop-top with electrons being accelerated near that region.  Radio observations combined with extreme-ultraviolet (EUV) imaging from the {\it Solar Dynamics Observatory} ({\it SDO})/Atmospheric Imaging Assembly (AIA, \citet{lemen_etal2012}) also reported that the hot loop-top and overlying regions are associated with coronal reconnection \citep{chen_b_etal2015, petrosian_2018, gary_etal2018}. The combination of these observations strongly suggest that hot loop-top sources (often having a cusp or candle flame morphology) and associated overlying structures are important for understanding the energy release process during a standard flare.

Solar flares are often associated with coronal mass ejections (CMEs) \citep{zhang_etal2001, zhang_etal2004, cheng_etal2010}. But, not all flares are associated with CMEs \citep{yashiro_etal2005}. Depending on the existence of CMEs, flares divided into the eruptive and confined flares \citep{svestka&cliver_1992, wang&zhang_2007}. The characteristics of eruptive flares have been well described by the standard flare model. The characteristics of confined flares are studied by comparison of the physical parameters between the eruptive and confined flares \citep{falconer_etal2006, kliem&torok_2006, cheng_etal2011, chen_h_etal2015, harra_etal2016, zhang_etal2017, li_t_etal2019}. Those studies suggest that the magnetic configuration, such as overlying field strength, decay index, and non-potentiality of an active region core, could be a determining factor of a confined flare.
 
On the other hand, flares produced by low (below the cusp) atmospheric reconnection have been reported using observations of X-ray jets, EUV jets, and H$\alpha$ surges \citep{hanaoka1997, ohyama_etal1997, shibata_1998, moore&sterling_2007, bain&fletcher_2009, zhang&ji_2014, panesar_etal2016, sterling_etal2017}. Recent multi-wavelength observations with higher spatial and temporal resolution reported a series of mini-filament eruptions and blowout-jets producing eruptive and confined flares \citep{shen_etal2012, wang&liu_2012, li_x_etal2015, lim_etal2016, panesar_etal2016b, hong_etal2017, panesar_etal2018, li_t_etal2018, shen_etal2019}.
Some of the flares were ejective due to secondary reconnection between the blowout jet eruption (or mini-filament) and overlying coronal field \citep{moore&sterling_2007, panesar_etal2016, sterling_etal2017}. Others were confined by the strong overlying fields and observed as the X-ray flaring arches/loops with H$\alpha$ surges or strong footpoint brightenings \citep{tang&moore_1982, martin&svestka_1988, svestka_etal1989, fontenla_etal1991, hanaoka1997, moore_etal1999, li_t_etal2018}. Those flaring arches are also associated with remote brightenings at the other footpoint, energized by the accelerated electrons and X-ray plasma traversing along the arch.  Combining with the magnetic field extrapolation, \citet{wang&liu_2012} and \citet{shen_etal2019} showed that the flares with the remote brightenings have magnetic fan-spine structures. Ergo, the magnetic field configuration could also be the determining factor between an ejective or confined eruption in flares driven by low atmospheric reconnection. 

Quantitative measures of flare plasma properties and their dynamics is important for understanding the differentiating characteristics between these two types of flares (i.e., confined versus ejective). EUV spectroscopic observations provide relevant and necessary diagnostic tools. Temperature and density variations in flares have been investigated using sensitive spectral line ratios or DEM analysis \citep{doschek_1999, mctiernan_etal1999, hara_etal2011, sun_etal2014, polito_etal2016a, warren_etal2018}. Flare plasma dynamics, such as the energy dissipation process and its response, are also traceable through the Doppler velocities derived from EUV spectral analysis. For example, spectroscopic observations have been used to study the inflow and outflow components of magnetic reconnection \citep{wang_etal2007, hara_etal2006, hara_etal2011, tian_etal2014}, as well as evaporative flows caused by energy dissipation in the low atmosphere \citep{brosius2003, brosius2009, brosius2013a, hara_etal2008_1, hara_etal2011, milligan&dennis_2009, tian_etal2015, polito_etal2015, polito_etal2016b, lee_etal2017}. Also, turbulent motions in flare plasma have been identified by the non-thermal motions that are measured by the excess of the line width \citep{hara_etal2006, doschek_etal2014, jeffrey_etal2016}. 

Since the launch of the {\it Hinode}/EUV Imaging Spectrometer (EIS, \citet{culhane_etal2007a}), which provides high spatial and spectral resolution across a wide temperature range, it is possible to resolve plasma properties in flare structures. For example, a comprehensive study of a flare observed on the solar disk was performed using EIS observations by \citet{hara_etal2011}. Figure~11 therein presents the flare structure with temperature and density distribution as well as flows (evaporation, reconnection inflow, and outflow) traced by the Doppler velocity of the different temperature lines. The study provides observational results of flare plasma properties and dynamic structures for a standard flare, but the disk observation made it difficult to distinguish the loop-top region and to explore the height distribution of flare plasma properties due to the superposition of plasma along the line of sight through the solar atmosphere. The height (spatial) distribution of the plasma properties, such as temperature, density, and flow velocities are critical parameters for understanding the energy transport process in the solar atmosphere.

Therefore, spectral observations of limb flares are needed to investigate the flare plasma properties with height distribution, even though the emission includes the superposition of several different loop structures along the line of sight. So far, however, there are few suitable spectroscopic observations of limb flares due to the difficulty of predicting nominal slit placement {\it a priori} \citep{hara_etal2006, doschek_etal2014}. Recently, \citet{warren_etal2018} and \citet{doschek_etal2018} analyzed an eruptive X8.3 
flare at the west limb, which includes a plasma sheet (indicative of the presence therein of a current sheet) above the post flare loop system. They measured the temperature, non-thermal velocity, and abundances of the loop and plasma sheet using EIS spectral measurements. To build upon these studies, additional spectroscopic observations and analysis of limb flares are necessary to map the height distribution of plasma properties and dynamics along flaring loops. 

We note that there is still a limitation of the limb observations, namely, that the structures are projected onto the plane of the sky. The measured Doppler velocity (line of sight velocity) is an average along the line of sight, and the direction of the flows depends on loop tilt angle. Therefore, to understand the configuration and the velocity structure of the flaring loop, we need to know the 3-D configuration of the loop. The {\it Solar~TErrestrial~RElations~Observatory} ({\it STEREO}, \citet{kaiser_etal2008}), which consists of two spacecraft, provides at least one additional vantage point. Combining {\it AIA} and {\it STEREO} observations, we are able to construct the 3-D flare loop. Similarly, \citet{imada_etal2013} investigated hot plasma motion using {\it Hinode}/EIS Doppler velocity measurements of high temperature lines at the limb combined with context observations from {\it STEREO's} EUV Imager (EUVI) to find the hot flows near the loop-top regions.  

To confirm the possibility that magnetic reconnection produces the flare, we also need to inspect the magnetic field topology in the corona. Typically, due to the lack of direct measurements in the corona, the coronal magnetic fields are extrapolated from photospheric vector magnetograms. It is well-known that the photosphere has a non-zero Lorentz force \citep{2001SoPh..203...71G}, which needs to be accounted for during the extrapolations. Within the framework of force-free extrapolations, a `pre-processing' of the photospheric data is often performed to minimize the Lorentz force in the vector magnetograms and provide a boundary that is compatible with the force-free condition \citep{2006SoPh..233..215W}. An alternative technique, used in this paper, is based on the use of non-force-free-fields (NFFFs), which are described by the double-curl Beltrami equation for the magnetic field {\bf B} \citep{2008SoPh..247...87H}. The numerical method for the non-force-free extrapolation for coronal magnetic fields are described in \citet{2008ApJ...679..848H,2010JASTP..72..219H}. Such extrapolations have been recently applied to model initial fields for flares and jets \citep{2018ApJ...860...96P,2018ApJ...869...69M}.

In the present study, we investigate a limb flare (SOL2014-01-13T21:51M1.3) exhibiting both characteristics of the standard flare (hot apparent cusp) and confined flare (flaring arch footpoint brightening but lacking CME). \citet{Hernandez_Perez_2019} recently reported complementary observations of this flare using {\it RHESSI} and {\it SDO}/AIA. We have comprehensively studied this flare using multi-wavelength imaging, spectroscopic and stereoscopic analysis, and magnetic field extrapolation. In particular, we estimate the plasma properties quantitatively using EUV spectroscopy and construct the 3-D configuration of the flare loop. Comparing these results with flare-loop plasma cooling models, we confirm that this flare is an example of the flaring arch \citep{tang&moore_1982, martin&svestka_1988, svestka_etal1989, fontenla_etal1991, hanaoka1997, moore_etal1999, li_t_etal2018}. The arch is mostly heated via low atmospheric reconnection, which is in contrast to the standard flare model energised by higher (i.e., supra-arcade) coronal reconnection.
 
We present the flare observations from each instrument in Section~\ref{sec:obs} and the observed parameters from imaging, spectroscopic, stereoscopic analysis, and magnetic topology in Section~\ref{sec:results}. We discuss the implication of the observed parameters in Section~\ref{sec:discussion}. A summary is given in Section~\ref{sec:summary}.

\section{Observational data} \label{sec:obs}

The M1.3 flaring event that occurred on 2014 January 13 was observed by multiple instruments, allowing us to examine its plasma properties and 3-D configuration in detail. {\it{SDO}}/AIA provides images of the full solar disk in seven EUV passbands (log ($T$) = 3.7$-$7.2\,K) at a high time cadence of 12\,s and a spatial resolution of 1.2$\arcsec$. The {\it Hinode}/X-ray Telescope (XRT, \citet{golub_etal2007}) captures higher temperature images (log ($T$) = 6.2$-$7.5\,K) with a cadence of 20$-$120\,s and a spatial resolution of $<$\,2.5$\arcsec$. The highest temperature imaging is supplied by {\it RHESSI} in SXR (12$-$20\,keV) and HXR (20$-$50\,keV) emission. We applied the ``Clean" method with 300 iterations and a temporal resolution of 5~minutes to increase the signal to noise ratio. The spatial resolution of {\it RHESSI} image is about 2$\arcsec$.

{\it Hinode}/EIS spectroscopically captured this limb flare with the study {\tt FlareResponse01}, which is a flare trigger study. This study ran from 21:50:44 to 22:44:12~UT consisting of 6~rasters, starting when the `hunter' study using the \ion{He}{2} emission detected strong intensity over the threshold. It is designed for observing flares using a moderate-cadence raster scan and large wavelength coverage. For that purpose, the 2$\arcsec$ slit scans 80 positions with sparse 3$\arcsec$ steps between them with a total field of view (FOV) of 239$\arcsec\times$304$\arcsec$. The exposure time at each position is 5~seconds, and each raster scan takes about 9~minutes. The study has 15 spectral windows covering the temperature range from log ($T$) = 4.9$-$7.2\,K (Table~\ref{tbl-1}). The hot plasma lines of \ion{Fe}{24}, \ion{Fe}{23}, and \ion{Ca}{17}, included in the wavelength window, are particularly useful for determining the flaring temperature plasma. The spectral resolution of EIS is about 0.022\,\AA. We use these measurements to investigate the flare plasma dynamics as well as the temperature and density distribution with height, which provides detailed quantitative plasma properties of the arch system.

\begin{deluxetable}{c c}
\tablecaption{List of the spectral windows in the EIS flare study of {\tt FlareResponse01}. \label{tbl-1}} 
\tablehead{\colhead{\hspace{2cm}Line ID (\AA)}\hspace{2cm} & \colhead{\hspace{2cm}$\log T_{max}$ (K)}\hspace{2cm} } 
	\startdata
	\ion{O}{4}	279.93	& 5.2 \\
	\ion{O}{5}	248.46	& 5.4 \\
	\ion{Si}{7}	275.36	& 5.8 \\
	\ion{Fe}{11} 188.22	& 6.2 \\
	\ion{Fe}{12} 186.89	& 6.2 \\
	\ion{Fe}{15} 284.16	& 6.4 \\
	\ion{S}{13} 256.69	& 6.5 \\
	\ion{Ca}{14} 193.87	& 6.7 \\
	\ion{Ca}{15} 181.90 & 6.7 \\
	\ion{Ca}{15} 200.97	& 6.7 \\
	\ion{Ca}{17} 192.85	& 6.8 \\
	\ion{Fe}{16} 262.98	& 6.8 \\
	\ion{Fe}{22} 253.17	& 7.1	 \\
	\ion{Fe}{23} 263.77	& 7.2 \\
	\ion{Fe}{24} 255.11	& 7.2 \\
	\enddata
	\tablecomments{The peak formation temperature of the spectral lines are taken from Chianti version 7.0 \citep{dere_etal1997, landi_etal2012}.}
\end{deluxetable}

We also utilize the SECCHI/EUVI telescope on {\it STEREO}-A to reconstruct the 3-D structure of the flare loop. SECCHI/EUVI provides 195 and 304\,\AA\ images of this flare from a different perspective. At the time of the observations, the separation angle between {\it STEREO}-A and Earth is about 130$^{\circ}$, therefore, {\it STEREO}-A observes the flare near the limb from the opposite direction. This large separation angle allows us to extrapolate the 3-D path of the loop structure above the limb. The {\it STEREO} images have a spatial resolution of 3.2$\arcsec$ and a cadence of 2.5~minutes.

In addition, we compared the observed coronal structure with the magnetic topology derived using vector magnetograms from {\it SDO}/HMI. The flare occurred at the west limb, so the vector magnetic field at the flaring time is not available. Instead, we use HMI vector magnetograms from January 11 at 23:00~UT, three days before the flare. The temporal and spatial resolution of the HMI vector magnetograms is about 12~minutes and 1$\arcsec$, respectively.

\section{Analysis and results} \label{sec:results}

\subsection{Overview of the M1.3 flare on 2014 January 13} \label{subsec: overview_obs}

\begin{figure}
\epsscale{0.9}
\plotone{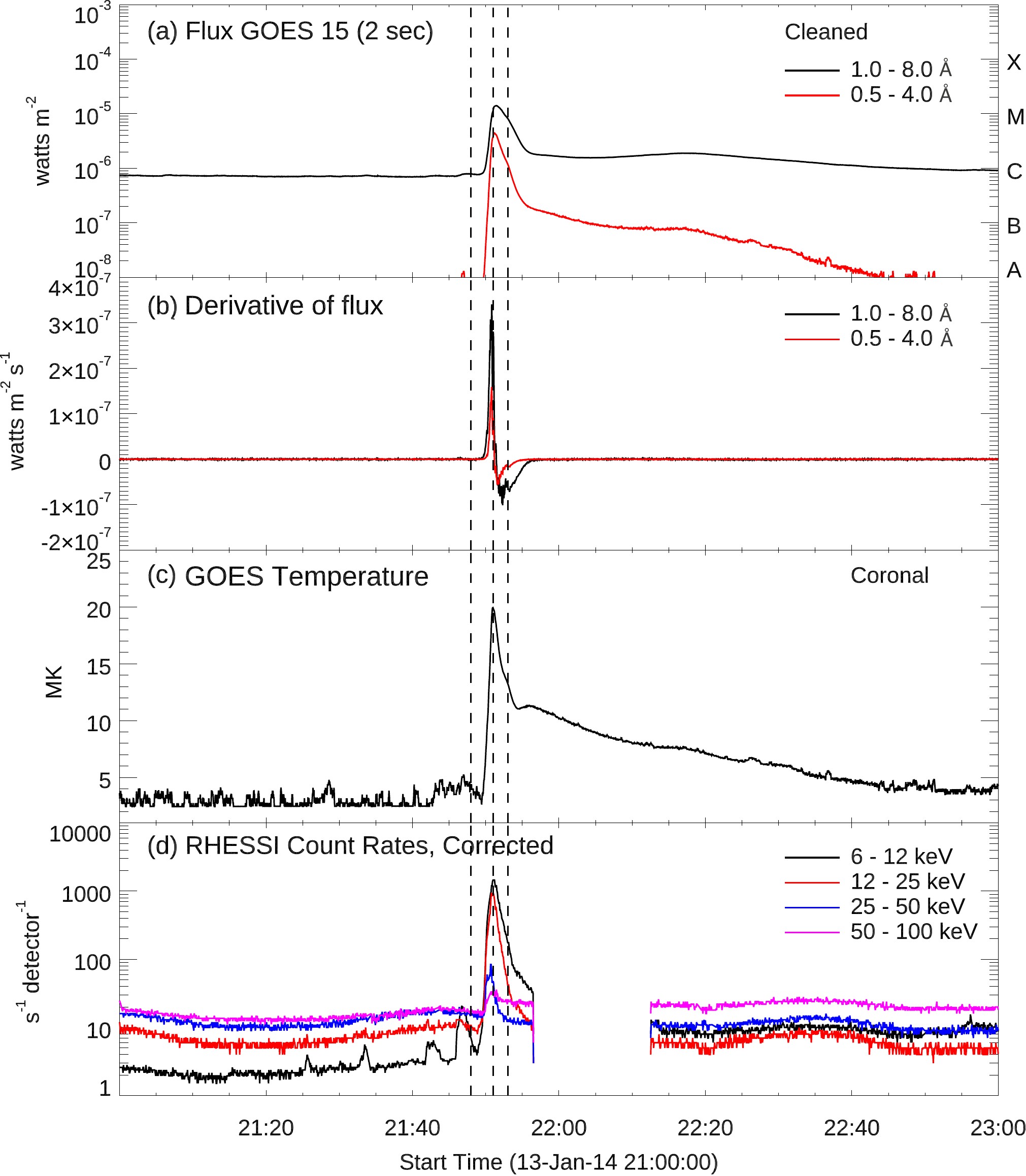} \caption{(a) {\it GOES} SXR light curves of the M1.3 flare on 2014 January 13 21:48~UT and (b) the time derivative of the SXR flux. (c) The temperature calculated from the  {\it GOES} flux ratio with coronal abundance. (d) The {\it RHESSI} count rates for the different energy bands. The vertical dashed lines indicate the flare start, peak, and end time of this flare in the {\it GOES} X-ray flare list.  \label{fig1}}
\end{figure}

We have investigated an M1.3 flare that occurred on 2014 January 13 near the solar limb. Figure~\ref{fig1} shows the {\it GOES} SXR (0.5$-$4\,\AA\ and 1$-$8\,\AA) light curves, the flux derivative, and the {\it{RHESSI}} hard (30$-$100\,keV) and soft (12$-$25\,keV) X-ray light curves of the flare event. The M-class flare started around 21:48~UT and peaked at 21:51~UT. The flare emission rapidly decayed to half its peak flux by about 21:53~UT, which shows that the flare heating was impulsive. 

\begin{figure}
\epsscale{0.9}
\plotone{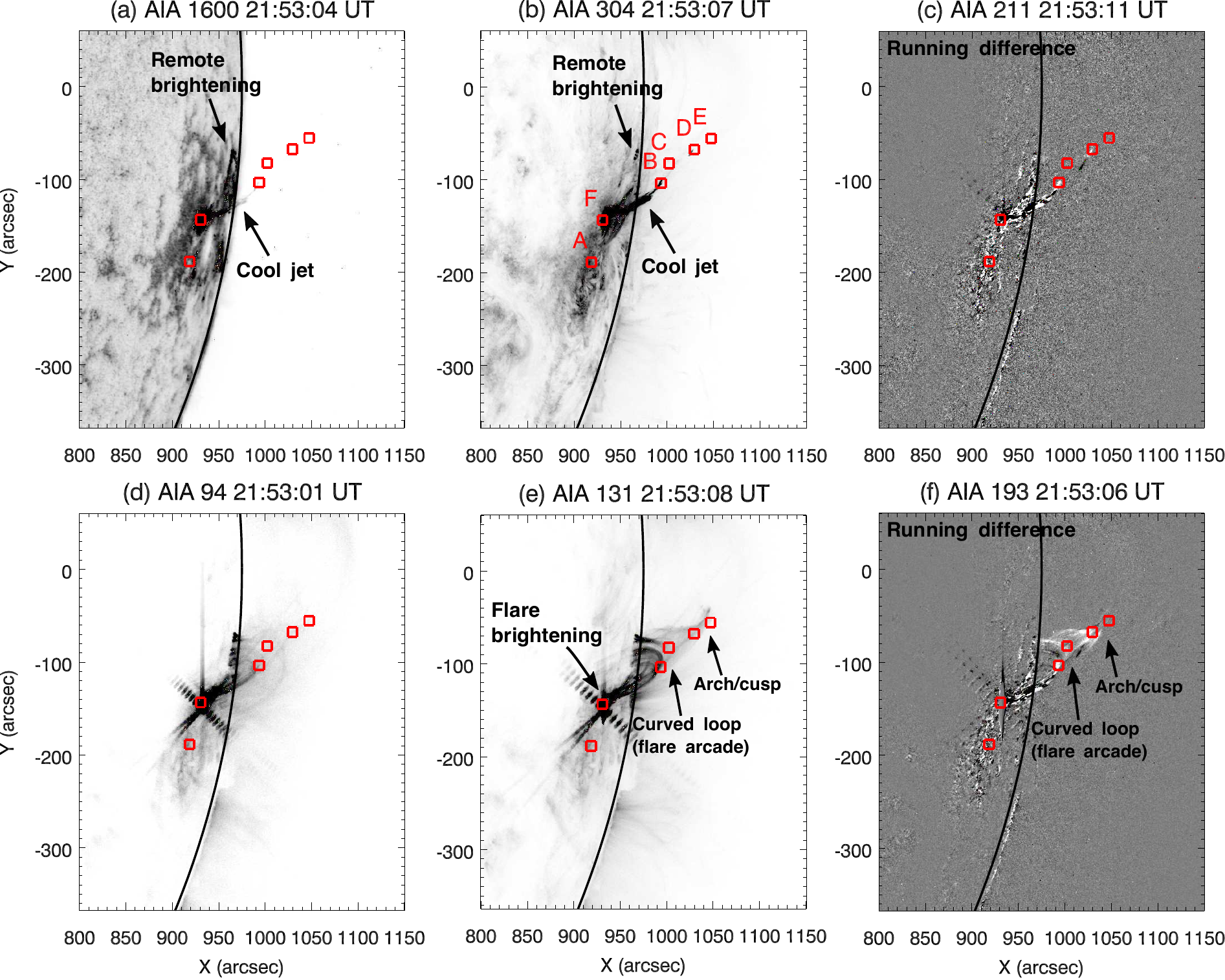}
\caption{Context {\it SDO}/AIA images for the M1.3 class flare on 2014 January 13 in different channels. Intensity scale is reversed in panels (a), (b), (d) and (e). Panels (c) and (f) show running difference images. The red boxes correspond to the locations for the light curves in Figure~\ref{fig5}. 
An animation of this figure is available. The locations for the light curves are marked with the blue boxes in panel (b) in the animation. The video begins after 21:40 UT and ends ater 22:20 UT. The realtime duration of the video is 18 seconds.   \label{fig2}}
\end{figure}

Figure~\ref{fig2} (and its associated animation) shows context images of the flare obtained by $\it{SDO}$/AIA at 21:53~UT. The AIA intensity images in different channels (i.e., 1600, 304, 211, 94, 131, and 193\,\AA) exhibit variable temperature-dependent plasma behavior. Their peak temperature responses are given in Table~\ref{tbl-2} \citep{lemen_etal2012, odwyer_etal2010}. The AIA 193\,\AA\ channel has temperature response peaks in both 1.5$-$2~MK and 17$-$20~MK ranges. Due to the bright background emission from 1.5$-$2~MK plasma, it is difficult to distinguish the enhancement of hot temperature (17$-$20~MK) emission in the 193\,\AA\ intensity images. Therefore, for the AIA 193\,\AA\ channel, we display the running difference image (panel (f)) in order to expose the increased emission at the cusp. 

For comparison, the running difference image of the AIA 211\,\AA\ channel (response peak of $\sim$2~MK) is shown in panel (c). The hot cusp is not observed in the 211\,\AA\ difference image, which confirms that the enhancement of the emission at the flaring arch/cusp region in 193\,\AA\ does not come from 1$-$2~MK plasma. The AIA multi-channel images and the corresponding movie depict (1) an apparent flare brightening at the lower eastern footpoint region at the flare start time (21:48~UT), and (2) a cool jet ejected in the AIA 1600 and 304\,\AA\ channels near the flare peak time ($\sim$21:51~UT). After the jet ejection, hot plasma along the loop expands upward and develops a flaring arch/cusp in the 94, 131, and 193\,\AA\ channels as the brief flare begins to decay ($\sim$21:53~UT). 

\begin{figure*}
\epsscale{1.1}
\plotone{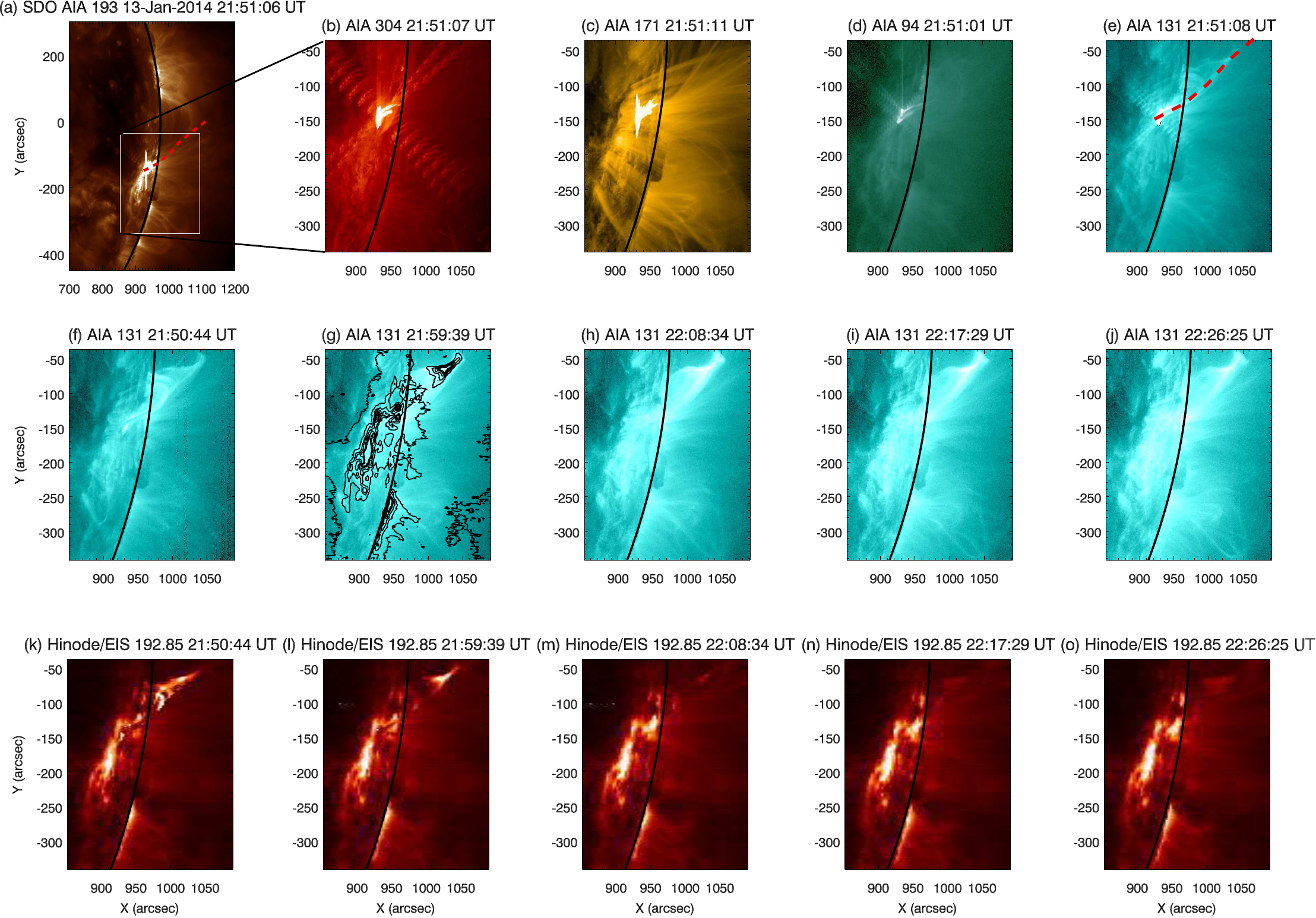} \caption{Aligned intensity images of {\it SDO}/AIA and {\it Hinode}/EIS 192\,\AA\ for the M1.3 class flare on 2014 January 13. (a)-(e): AIA 193, 304, 171, 94, and 131\,\AA\ channel images taken near the flare start time ($\sim$ 21:51 UT). The white box in panel (a) indicates the FOV of EIS scan raster. (f)-(j): Pseudo raster scan AIA 131\,\AA\ images taken at each EIS scan slit time. The EIS 192\,\AA\ intensity contour is overlaid in panel (g). (k)-(o): EIS 192\,\AA\ total intensity images during the flare. The red dashed lines in panels (a) and (e) correspond to the slit location for the time-distance diagrams in Figure~\ref{fig6}. \label{fig3}}
\end{figure*}

Figure~\ref{fig3} shows the aligned intensity images of $\it{SDO}$/AIA and $\it{Hinode}$/EIS. The top panels (a$-$e) are the AIA intensity images in different channels (i.e., 193, 304, 171, 94, and 131\,\AA) with same FOV of the {\it Hinode}/EIS raster at the flare peak time (21:51~UT). The middle row displays the pseudo raster image using the  AIA 131\,\AA\ channel images observed at the nearest time of the {\it Hinode}/EIS raster scan for comparison (panels f$-$j). The bottom panels (k$-$o) show the {\it Hinode}/EIS total intensity image for the \ion{Ca}{17}~192.85\,\AA\ wavelength window, which includes \ion{Fe}{11}, \ion{}{12}, \ion{}{14}, \ion{}{24}, \ion{O}{5}, and \ion{Ca}{17} emission. The marked timings indicate the start time of the raster scan. The intensity contour of the EIS \ion{Ca}{17}~192.85\,\AA\ emission is overlaid onto the AIA 131\,\AA\ image in the middle row (panel g).

\begin{deluxetable}{c c c c c }
\tablecaption{Intensity peak times of the flare plasma in the different instruments and wavelength channels, time delays relative to the flare peak time from {\it GOES}, and the peak response temperature for each channel. } \label{tbl-2} 
\tablehead{\colhead{Instrument} & \colhead{Observed\tablenotemark{a}}& \colhead{Peak response\tablenotemark{b}} & \multicolumn{2}{c}{Peak time [UT]}  \\ 
\colhead{Wavelength} & \colhead{spectra} & \colhead{temperature [MK]} & \multicolumn{2}{c}{(Time delay [s])} \\
\cline{4-5} 
 \colhead{} & \colhead{} & \colhead{} & \colhead{Region C} & \colhead{Region D} }
 	\startdata
	GOES 0.5$-$4\,\AA\ (1.0$-$8 \AA) & \nodata & 19.9 & 21:50:57 (0) & 21:50:57 (0) \\
	XRT Be$\_$Thick & \nodata & 12.6 & 21:53:18 (141) & 21:53:18 (141) \\
	AIA 193	& {\ion{Fe}{24} (\ion{Fe}{12})}  & {17.8 (1.6)} & {21:53:54 (177)} & 21:56:42 (345) \\
	AIA 131	& {\ion{Fe}{21} (\ion{Fe}{8})}  & {11.2 (0.5)} & {21:59:56 (539)} & 22:12:56 (1319) \\
	AIA 94	& {\ion{Fe}{18} (\ion{Fe}{10})}  & {7.1 (1.1)} & {22:19:01 (1655)} & 22:37:25 (2788) \\
	{AIA 335}	& {\ion{Fe}{16} (\ion{Mg}{8})}  & {2.8 (0.79)} & \nodata & \nodata \\
	{AIA 211}	& {\ion{Fe}{14}}  & {2} & \nodata & \nodata \\
	{AIA 171}	& {\ion{Fe}{9}}  & {0.8} & \nodata & \nodata \\
	{AIA 304}	& {\ion{He}{2}}  & {0.08} & \nodata & \nodata \\
	\enddata
	\tablecomments{The intensity peak times from the XRT and AIA channels determined from the  flare loop and cusp regions (labels C and D in Figure~\ref{fig5}).}
	\tablenotetext{a}{{The dominant spectral lines for the AIA channels \citep{odwyer_etal2010, lemen_etal2012}. The AIA 193, 131, 94, and 335\,\AA\ passbands have two peaks in hot and cool temperatures. We assume that the emission comes from the hot temperature lines during the flare. Cool temperature lines are given in parenthesis.}}
	\tablenotetext{b}{{The temperature for {\it GOES} is determined from the {\it GOES} short and long emission ratio. XRT peak response temperature is referred from \citet{narukage_etal2014_xrtcal} and \citet{odwyer_etal2014}. }} 
\end{deluxetable}

\clearpage

\begin{figure}
\epsscale{1.2}
\plotone{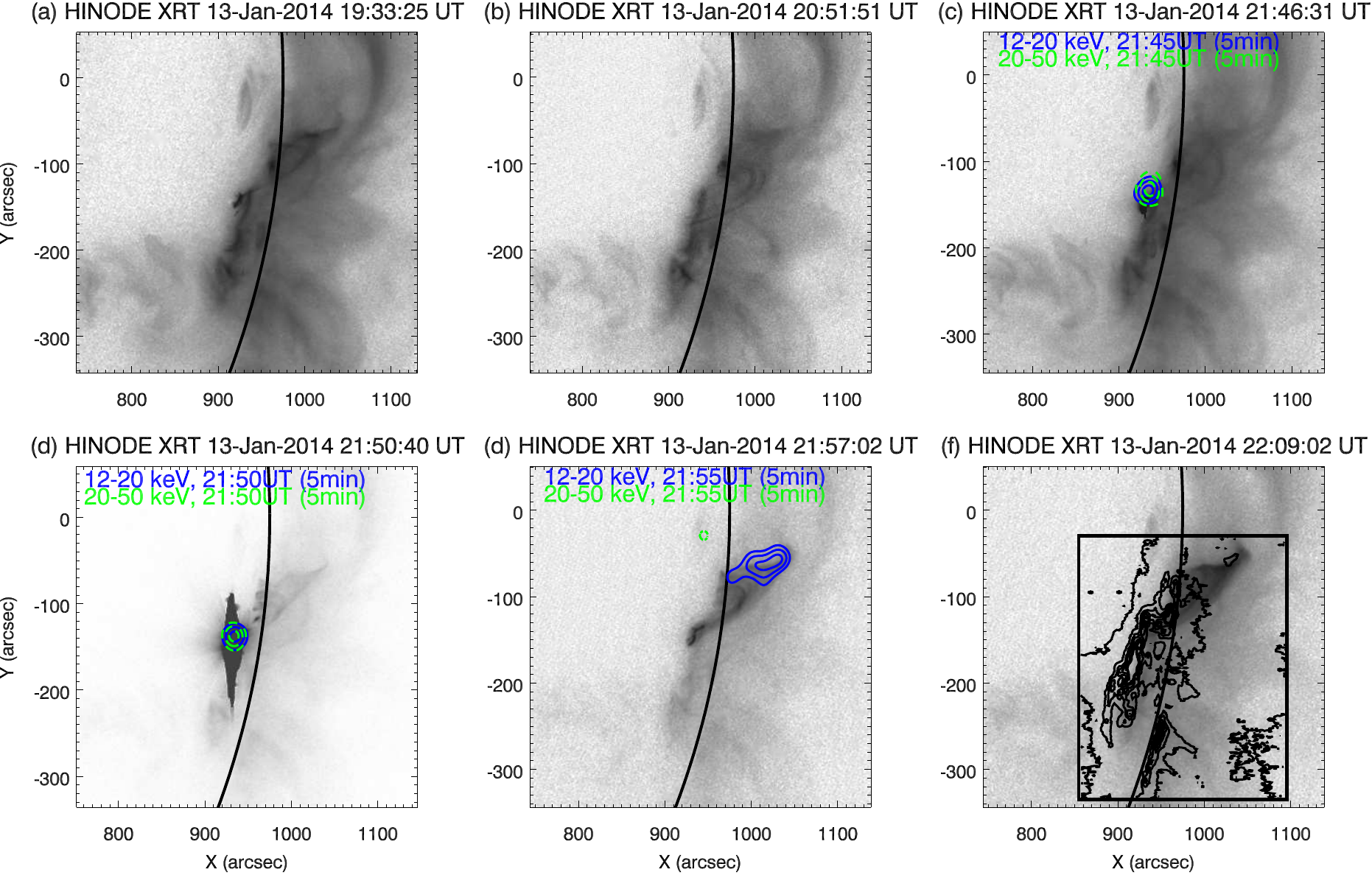} \caption{Reverse scale {\it Hinode}/XRT Be\_thin filter images before (a: $\sim$19:33~UT) and after the flare peak (f: $\sim$22:09~UT) with HXR (green dashed: 20$-$50\,keV) and SXR (blue: 12$-$20\,keV) contours overlaid from {\it RHESSI} cleaned images covering 21:45, 21:50, and 21:55~UT. The white box and contours in panel (f) correspond to the {\it Hinode}/EIS FOV and 192\,\AA\ intensity contours, respectively.  \label{fig4}}
\end{figure}

Figure~\ref{fig4} shows the {\it Hinode}/XRT Be\_thin images before ($\sim$19:33~UT) and after ($\sim$22:09~UT) the flare peak at higher temperatures. Panels (d) and (e) show the XRT intensity images with the cleaned {\it RHESSI} HXR and SXR intensity contours (50, 60, 70, 80, and 90\%) overlaid. The colored contours indicate HXR (20-50\,keV: orange and purple) and SXR (12-20\,keV: green and blue) emission observed by {\it RHESSI} at the closest time to the XRT images, respectively.

The XRT images reveal that the hot plasma fan loop/arch already exists before the flare onset (Figure~\ref{fig4}, a$-$c). At the {\it GOES} flare start time ($\sim$21:48~UT), a strong footpoint brightening was also detected in the XRT images, and the cusp shape loop structure intensity enhanced with time.  The temporal behavior in the HXR (20$-$50\,keV) and SXR (12$-$20\,keV) emission is supplied by the {\it RHESSI} observations (the contours in the (d) and (e) panels). The HXR emission only appears early in the event between 21:45 and 21:54~UT near the footpoint brightening, and the SXR emission moves from the footpoint to the loop top region from 21:50 to 21:59~UT.

\subsection{Imaging analysis} \label{subsec:imaging}

\begin{figure*}
\epsscale{1.1}
\plotone{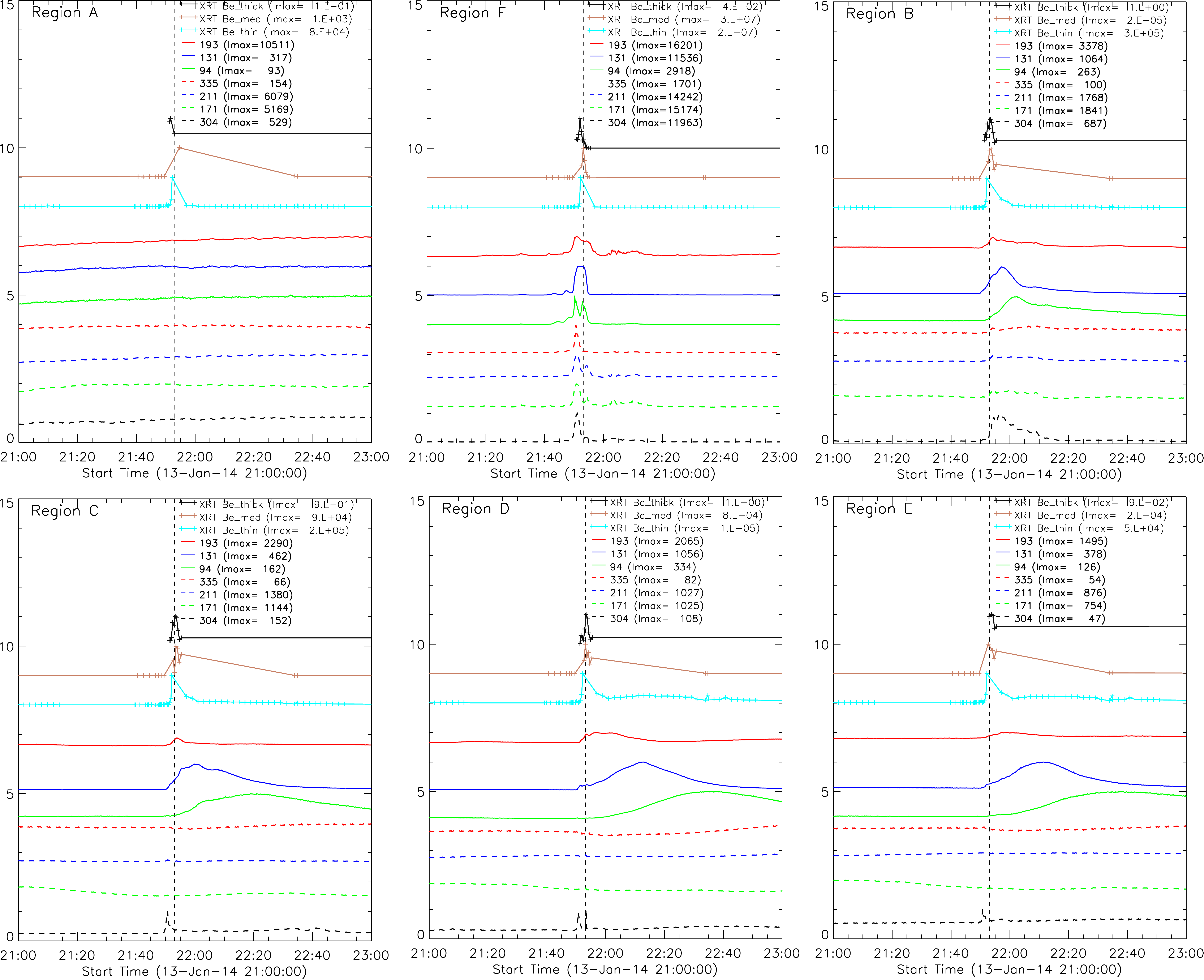}\caption{Light curves in 7 AIA EUV channels and 3 XRT SXR filters at different locations along the loop. These locations (A$-$F) are marked with red boxes in Figure~\ref{fig2}. The vertical dashed lines indicate the observing time of AIA images in Figure~\ref{fig2}. \label{fig5}}
\end{figure*}

Using the multi-wavelength imaging observations from EUV to X-ray, we have investigated the temporal behavior of the plasma in the flaring fan loop at different temperatures. Figure~\ref{fig5} shows 7 EUV and 3 SXR light curves from the footpoint region (F) and along the flaring loop (B$-$E). The light curve from the region A shows the active region emission near the flare as a reference point. From top to bottom, we plotted the light curves in descending temperature order. As previously discussed, the response of the AIA~193\,\AA\ channel is double-peaked at $\log T\sim$1.5$-$2~MK and $\sim$17$-$20~MK \citep{lemen_etal2012}. Since the EIS measurements conclude that the loop-top is dominated by hot \ion{Fe}{24} emission, we assume that the prevailing AIA~193\,\AA\ channel peak response temperature is $\sim$18~MK in the cusp region. 

The relatively cooler coronal emission (304, 171, 211, and 335\,\AA) is only observed at the flaring footpoint (region F) with a sharp peak at the start time of the flare. Meanwhile, the flaring temperature plasma of XRT and AIA 193, 131, and 94\,\AA\ emission, along the flaring loop/arch (region B$-$E) persists for a longer duration. The peaks in the AIA 304\,\AA\ light curve from the region C$-$E are due to the saturation pattern of the flaring footpoint brightening. Considering the temperature responses of AIA and XRT, the light curves show that the intensity peak of the loop top region has a time shift from the higher to lower temperature channels (i.e., XRT Be\_thick to AIA 94\,\AA) in descending temperature order. The time shift of the intensity peak is interpreted as the cooling of the plasma with height in the loop \citep{viall&klimchuk_2012, ryan_etal2013, thiemann_etal2017}.

\clearpage

\begin{figure}
\epsscale{0.7}
\plotone{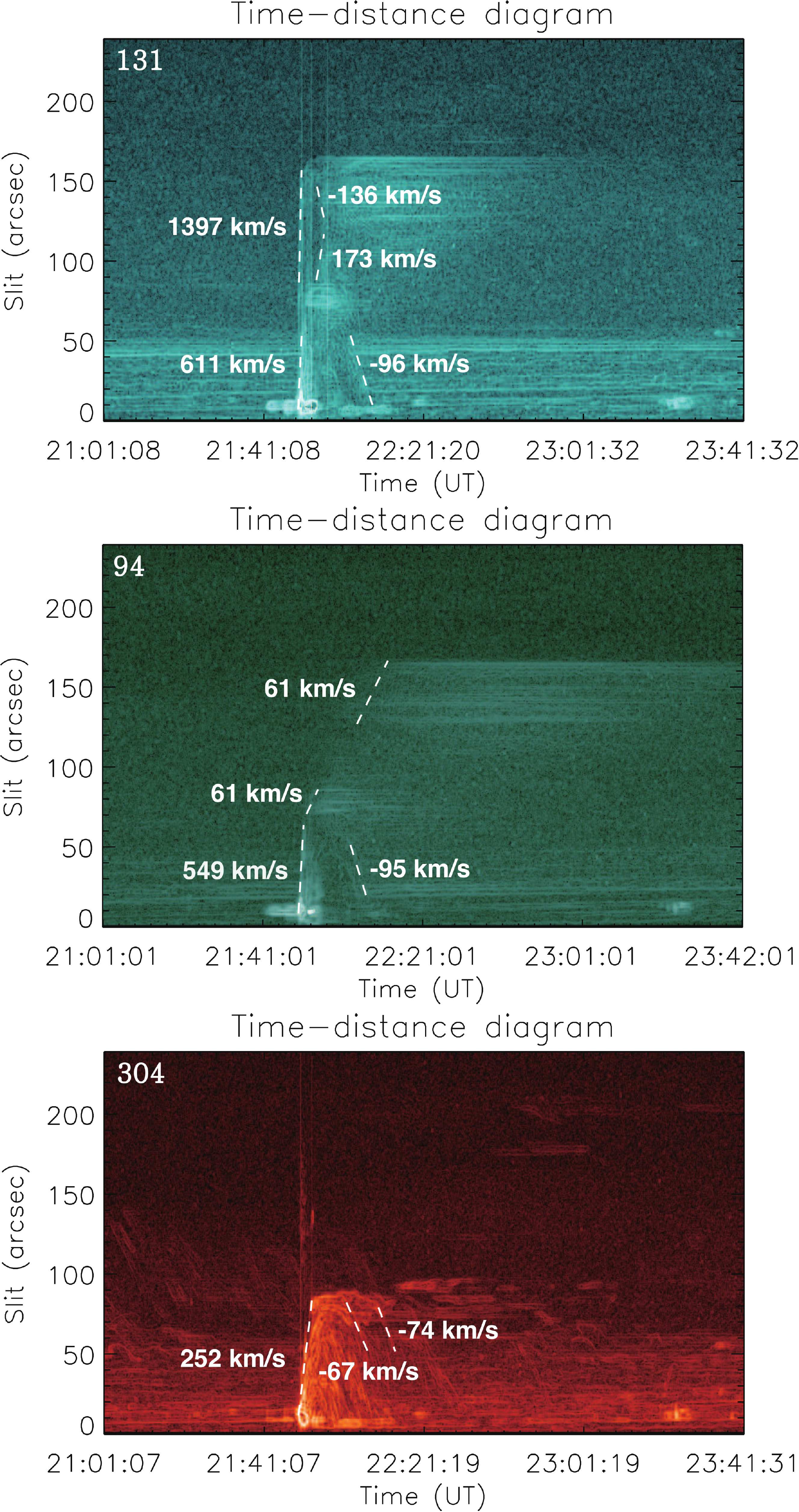} \caption{Time-distance diagrams from a virtual slit along the loop direction (dashed line in Figure~\ref{fig3}\,(e)) for the AIA 131, 94, and 304\,\AA\ channels. The white dashed lines tracing the flows are used for the velocity calculation. The measured velocities are listed beside the each  of lines. \label{fig6}}
\end{figure}

Flows are also detected as intensity peaks traveling along the arch system. Figure~\ref{fig6} depicts the time-distance diagrams illustrating these flows in the AIA 304, 94, and 131\,\AA\ passbands. In the AIA 304\,\AA\ channel, a cool jet structure is ejected at 21:48~UT (the flare start time), and its projected velocity on the plane of the sky is $\sim$250\,km\,s$^{-1}$. Then, the plasma descends with a velocity of $\sim$65$-$75\,km\,s$^{-1}$. \cite{panesar_etal2016} reported active region jet ejections with flares using AIA 304 and found that their speeds ranged 110$-$400\,km\,s$^{-1}$, which is consistent with this jet speed. The 131\,\AA\ diagram displays that expansion of coronal plasma outwards after the jet ejection with a velocity of $\sim$700\,km\,s$^{-1}$, and then the emission is sustained about an hour locally. The 94\,\AA\ intensity also shows an enhancement at different loop locations, but this emission does not appear to increase at the ejection site. These observations suggest that the cooling time varies spatially along the fan loop. 

\clearpage

\subsection{Spectroscopy} \label{subsec: spectroscopy}

\subsubsection{Line identification} \label{subsubsec: line_int}

\begin{figure*}
\epsscale{1.2}
\plotone{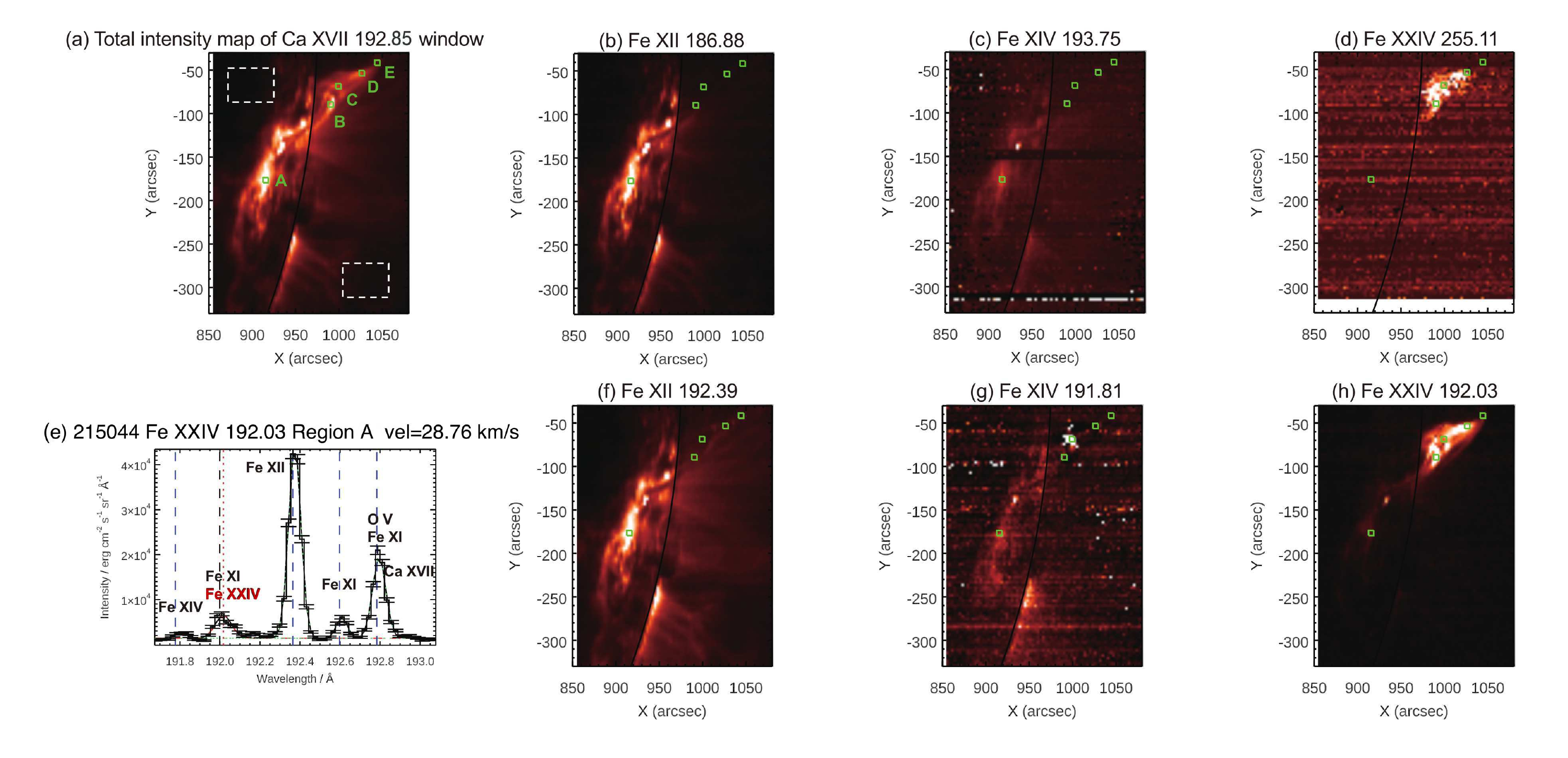} \caption{EIS intensity images from the EIS \ion{Ca}{17} 192.85\,\AA\ window at 21:50:44~UT (impulsive phase). (a) Total intensity image of the EIS \ion{Ca}{17} 192.85\,\AA\ window. The white dashed boxes correspond to the quieter regions used for measuring reference wavelengths. (b)$-$(d): EIS intensity images centered on the \ion{Fe}{12} 186.88, \ion{Fe}{14} 193.75, and \ion{Fe}{24} 255.11\,\AA\ lines derived from the each different wavelength window. The green boxes indicate the locations corresponding to the spectral line profiles. Panel (e) shows an example of the spectral line profiles in the EIS \ion{Ca}{17} 192.85\,\AA\ window for region A. The line identifications are marked for each spectral line. The red dotted and black dashed lines indicate the reference wavelength and fitted line center wavelength of \ion{Fe}{24} 192.03\,\AA, respectively. The blue dashed vertical lines represent the reference wavelengths of different spectral lines in the same wavelength window, \ion{Fe}{14}, \ion{Fe}{12}, \ion{Fe}{11}, and \ion{Ca}{17}. (f)$-$(h): EIS intensity images derived from multiple Gaussian fittings of spectral lines in the \ion{Ca}{17} 192.85\,\AA\ window for \ion{Fe}{12} 192.39, \ion{Fe}{14} 191.81, and \ion{Fe}{24} 192.03\,\AA, respectively. \label{fig7}}
\end{figure*}

\begin{figure}
\epsscale{0.9}
\plotone{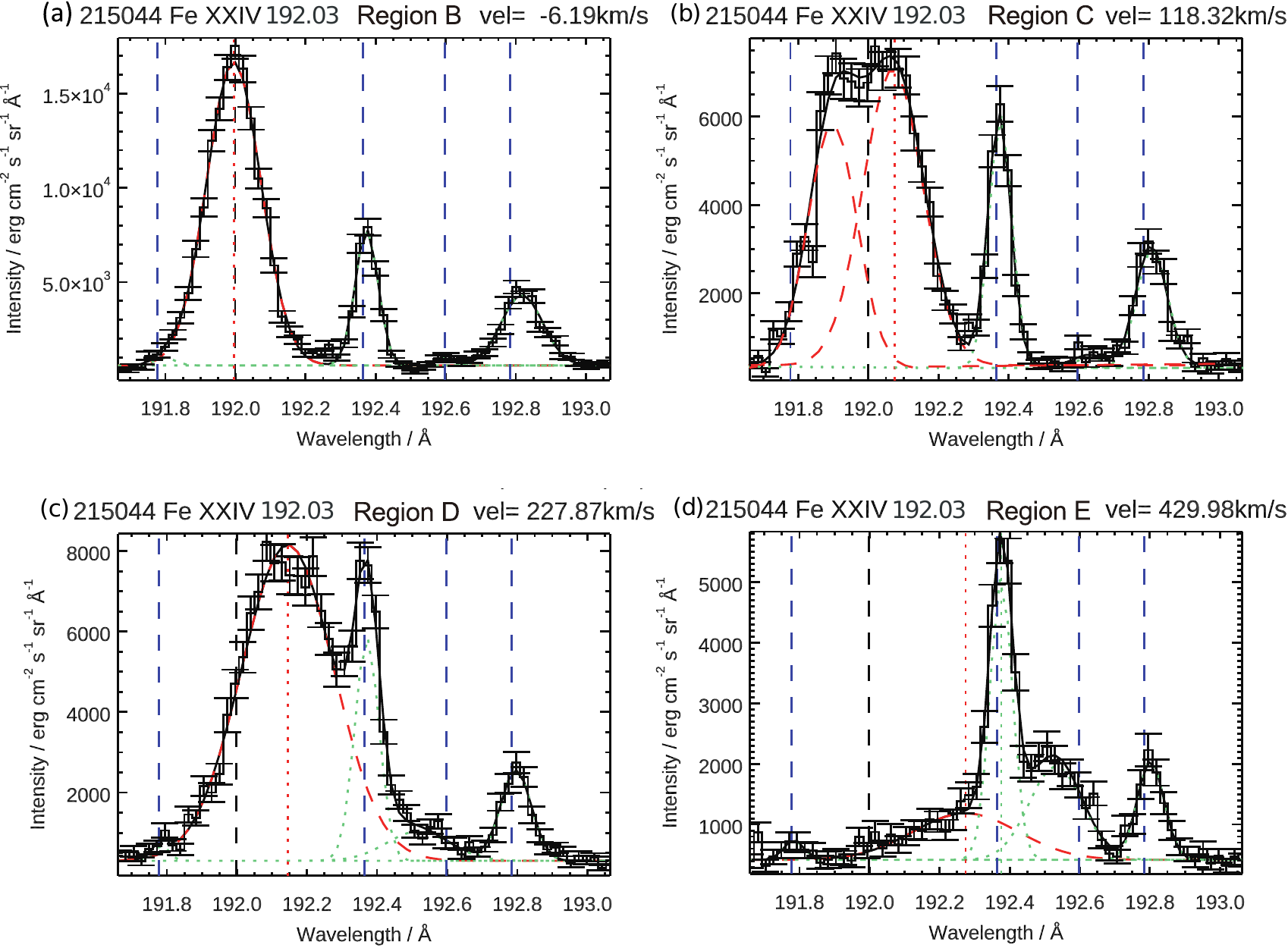} \caption{EIS spectral profiles from the EIS \ion{Ca}{17} 192.85\,\AA\ window at 21:50:44~UT (impulsive phase) for the locations marked in Figures~\ref{fig2} and \ref{fig7}. The black dashed vertical lines represent the reference wavelength of \ion{Fe}{24} 192.03\,\AA. The blue dashed vertical lines correspond to the reference wavelengths of different spectral lines in the same wavelength window, \ion{Fe}{14}, \ion{Fe}{12}, \ion{Fe}{11}, and \ion{Ca}{17}. The red dashed (\ion{Fe}{24}) and green dotted (\ion{Fe}{14}, \ion{Fe}{12}, \ion{Fe}{11}, and \ion{Ca}{17}) lines represent the fitted components of the multiple-Gaussian fitting analysis, respectively. The solid lines represents the fitted total line spectral profile. The red dotted vertical lines correspond to the center wavelength of the fitted component of \ion{Fe}{24}. \label{fig8}}
\end{figure}

\begin{figure*}
\epsscale{1.2}
\plotone{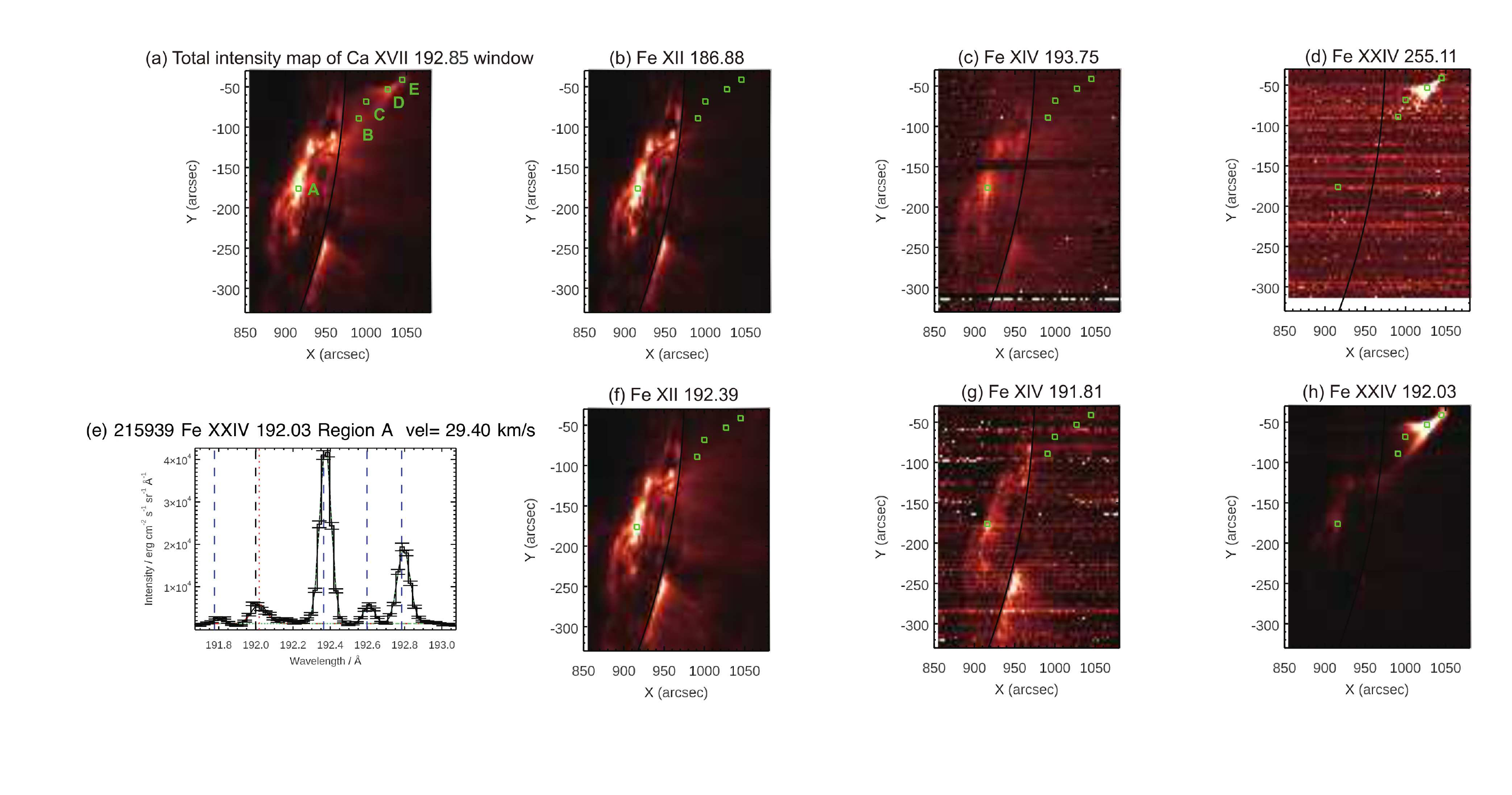} \caption{EIS intensity images from the EIS \ion{Ca}{17} 192.85, \ion{Fe}{12} 186.88, \ion{Fe}{14} 193.75, and \ion{Fe}{24} 255.11\,\AA\ windows at 21:59:39~UT (gradual phase). The format is the same as Figure~\ref{fig7}. \label{fig9}}
\end{figure*}

\begin{figure}
\epsscale{0.9}
\plotone{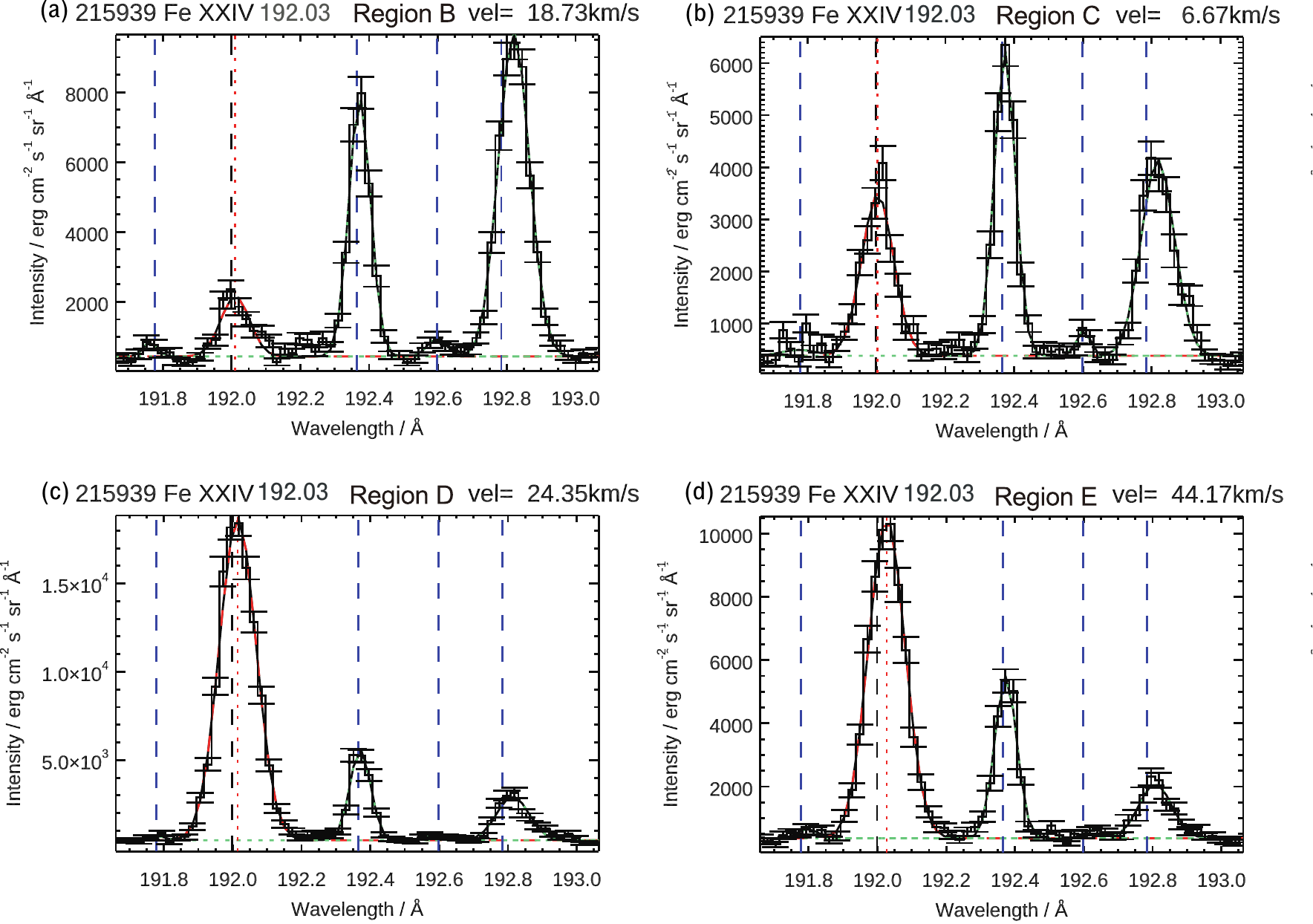} \caption{EIS spectral profiles from the EIS \ion{Ca}{17} 192.85\,\AA\ window at 21:59:39~UT (gradual phase) for the locations marked in Figures~\ref{fig2} and \ref{fig7}. The format is the same as Figure~\ref{fig8}. \label{fig10}}
\end{figure}

To obtain the intensity, Doppler velocity, and line width of the limb flare, we fitted the spectral lines that are listed in Table~\ref{tbl-1} using single and multiple Gaussians. Figures~\ref{fig7} and \ref{fig9} show examples of the intensity images from selected EIS spectral scans (start times of 21:50:44 and 21:59:39~UT, respectively). Figures~\ref{fig8} and \ref{fig10} present the spectral line profile of \ion{Fe}{24} (192.03\,\AA) from the EIS \ion{Ca}{17} 192.85\,\AA\ window. To obtain reference wavelengths for the EIS spectra, we made reference measurements in the quieter regions within the FOV, one from on disk and the other from at the limb (dashed boxes in Figure~\ref{fig7} (a)). Those reference wavelengths are marked with vertical dashed lines in the spectra shown in Figures~\ref{fig8} and \ref{fig10}. 

To identify the flaring temperature of the \ion{Fe}{24} 192.03\,\AA\ spectral line, which is blended with \ion{Fe}{11} (192.02\,\AA) and close to \ion{Fe}{12} (192.39\,\AA) and \ion{Fe}{14} (191.81\,\AA), we deconvolved the line intensities using similar temperature line intensities that are isolated in different wavelength windows. Figure~\ref{fig7} shows the EIS raster scan at the beginning of the flare.  Panel (a) shows the total intensity image of the \ion{Ca}{17} 192.85\,\AA\ window, which includes the wavelengths from 191.67 to 193.08\,\AA. Panels (f)$-$(h) show the intensity images derived from multiple-Gaussian fits to the spectra in panel (a) for \ion{Fe}{12}, \ion{Fe}{14}, and \ion{Fe}{24}, respectively. Panels (b)$-$(d) give the intensity images from the same ionization states of the iron lines provided in panels (f)$-$(h), but centered on different wavelengths (i.e., \ion{Fe}{12} 186.88, \ion{Fe}{14} 193.75, and \ion{Fe}{24} 255.11\,\AA). These lines are well isolated and not blended with other lines. The line intensities emitted from similar temperature plasma would produce similar intensity images. From the comparison, we could identify the hot flaring component at the loop-top source, seen only in the \ion{Fe}{24} intensity images and not in the other isolated \ion{Fe}{12} (Figure~\ref{fig7}~(b)) and \ion{Fe}{14} (Figure~\ref{fig7}~(c)) intensity images. Also, the footpoints and the bottom legs of the active region (box A) do not produce detectable emission from \ion{Fe}{24}. Therefore, we have identified \ion{Fe}{24} emission in Figures~\ref{fig8} and \ref{fig10} along the loop structure and measured the Doppler velocity along the line of sight. 

\clearpage

\subsubsection{Doppler velocity} \label{subsubsec:doppler_vel}

\begin{figure}
\epsscale{0.8}
\plotone{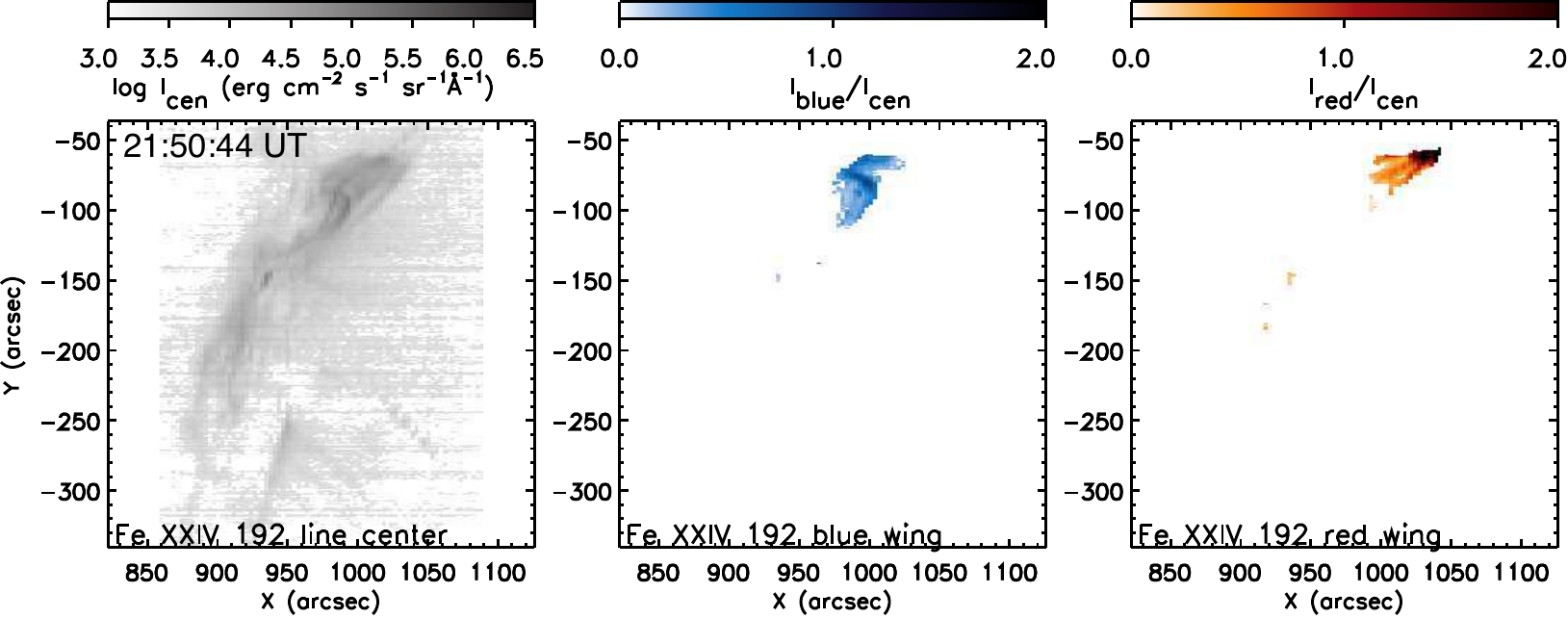}
\plotone{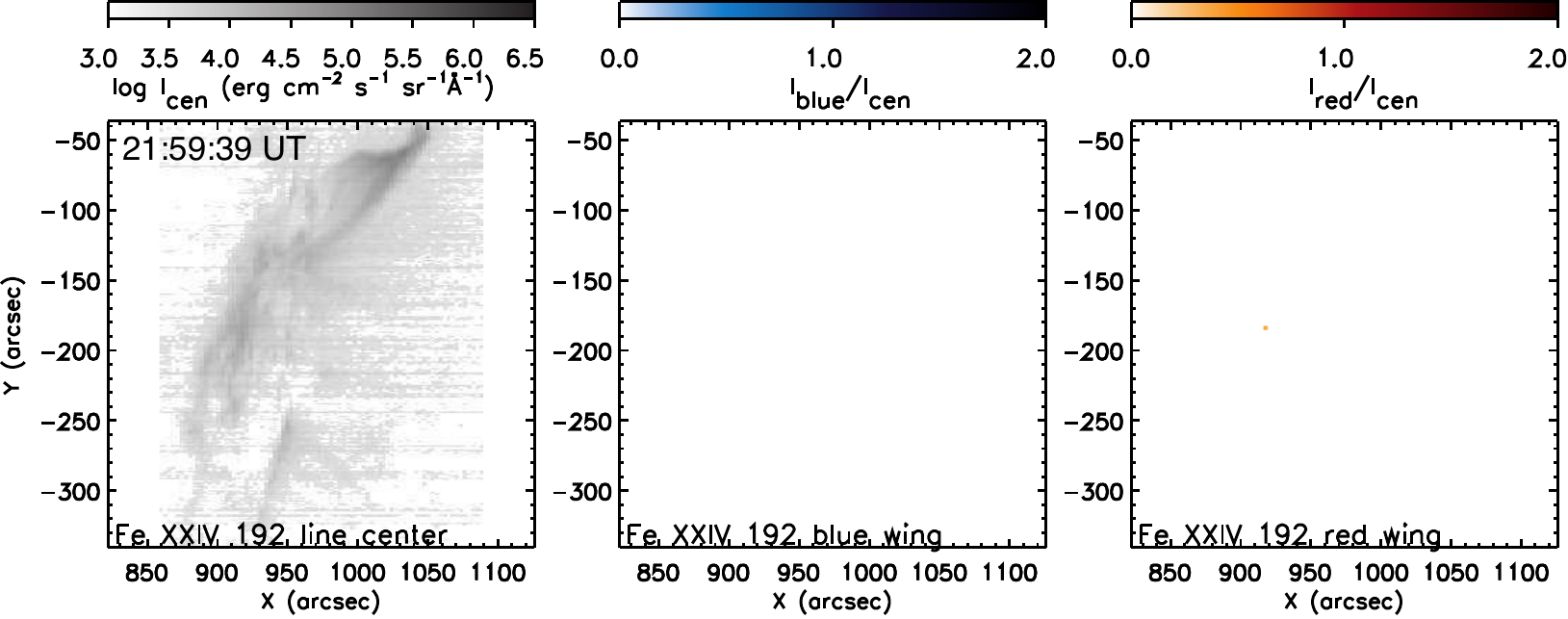} 
\caption{Intensity (left) and Doppler velocity maps (middle: blue-shifted, right: red-shifted) of \ion{Fe}{24} 192.03\,\AA\ from the wing and center intensity ratio for impulsive phase (top) and gradual phase (bottom). \label{fig11}}
\end{figure}

Figure~\ref{fig8} shows the spectral line profiles of \ion{Fe}{24} (192.03\,\AA) at the beginning of the flare, and we find strong redshifts ($\sim$300$-$500\,km\,s$^{-1}$) along the loop-top region (locations D and E in Figures~\ref{fig7} and \ref{fig8}). Also several tens of arcseconds below, the small curved loop structure has a weak blue-shifted emission in \ion{Fe}{24}. Figure~\ref{fig11} displays the \ion{Fe}{24} 192.03\,\AA\ line center and wing intensity ratio images (shifted from the center by 100\,km\,s$^{-1}$), which imply the blue- and red- shifted emission map. The first raster (upper panel) observed at the flare start depicts the velocity distribution clearly -- strong red-shift at the loop-top region and weak blue-shift at the curved loop structure. The second raster (bottom panel) observed at 21:59~UT shows that all of the shifted emission vanishes and the intensity of the hot plasma component is enhanced. 

Due to the projection effect of the limb event, we need to know the 3-D loop trajectory in order to derive the flow direction along the loop (Section~\ref{subsec:stereoscopy}). We discuss the strongly red-shifted hot plasma with respect to the 3-D loop configuration in Section~\ref{subsec:plasmaflows}.

\clearpage

\subsubsection{Temperature} \label{subsubsec:temp}

\begin{figure}
\epsscale{0.9}
\plotone{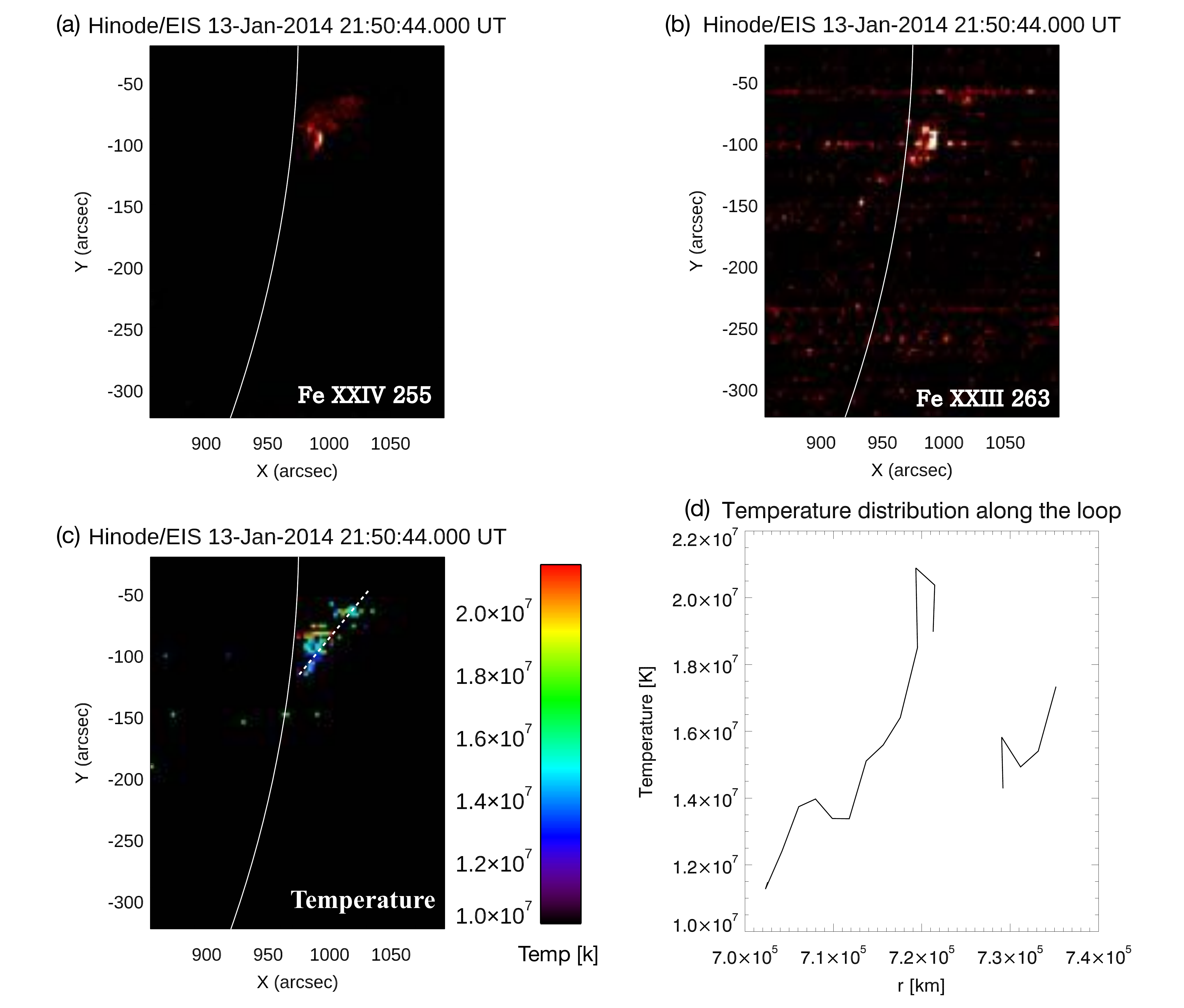} \caption{Temperature map (panel (c)) from intensity ratio of \ion{Fe}{24} 255.1\,\AA\ (panel (a)) and \ion{Fe}{23} 263\,\AA\ (panel (b)). Panel (d) shows the temperature distribution along the loop height, which is marked by the white dashed line in panel (c). \label{fig12}}
\end{figure}

To confirm the hot loop-top source and the temperature distribution along the loop, we measured the temperature using the line ratio from \ion{Fe}{24} (255.1\,\AA) and \ion{Fe}{23} (263.8\,\AA) \citep{hara_etal2011, warren_etal2018}. This ratio provides the plasma temperature in the range of $\log T = 6.5 - 7.5 $. Figure~\ref{fig12} shows each intensity map of those spectral lines (panels (a) and (b)) and the temperature map determined by the intensity ratio from the raster during the impulsive phase starting at 21:50~UT (panel (c)).  From this analysis, we confirm that the loop-top emission comes from hot plasma having temperature over 10\,MK and up to 22\,MK. Panel (d) of Figure~\ref{fig12} shows the temperature distribution with height in the loop. We estimated the temperature gradient with height in the loop by fitting the temperature distribution. The resulting gradient shows that the curved loop-top region has the highest temperature plasma and the temperature increases with height. The plasma above the loop-top region also shows high temperature plasma above 15\,MK, also increasing with height, but lower in temperature than the loop top region. Between these two structures, the intensities of the two spectral lines are so weak that we can not measure the temperature. 

\clearpage

\subsubsection{Density} \label{subsubsec:dens}

\begin{figure}
    \epsscale{0.9}
    \plotone{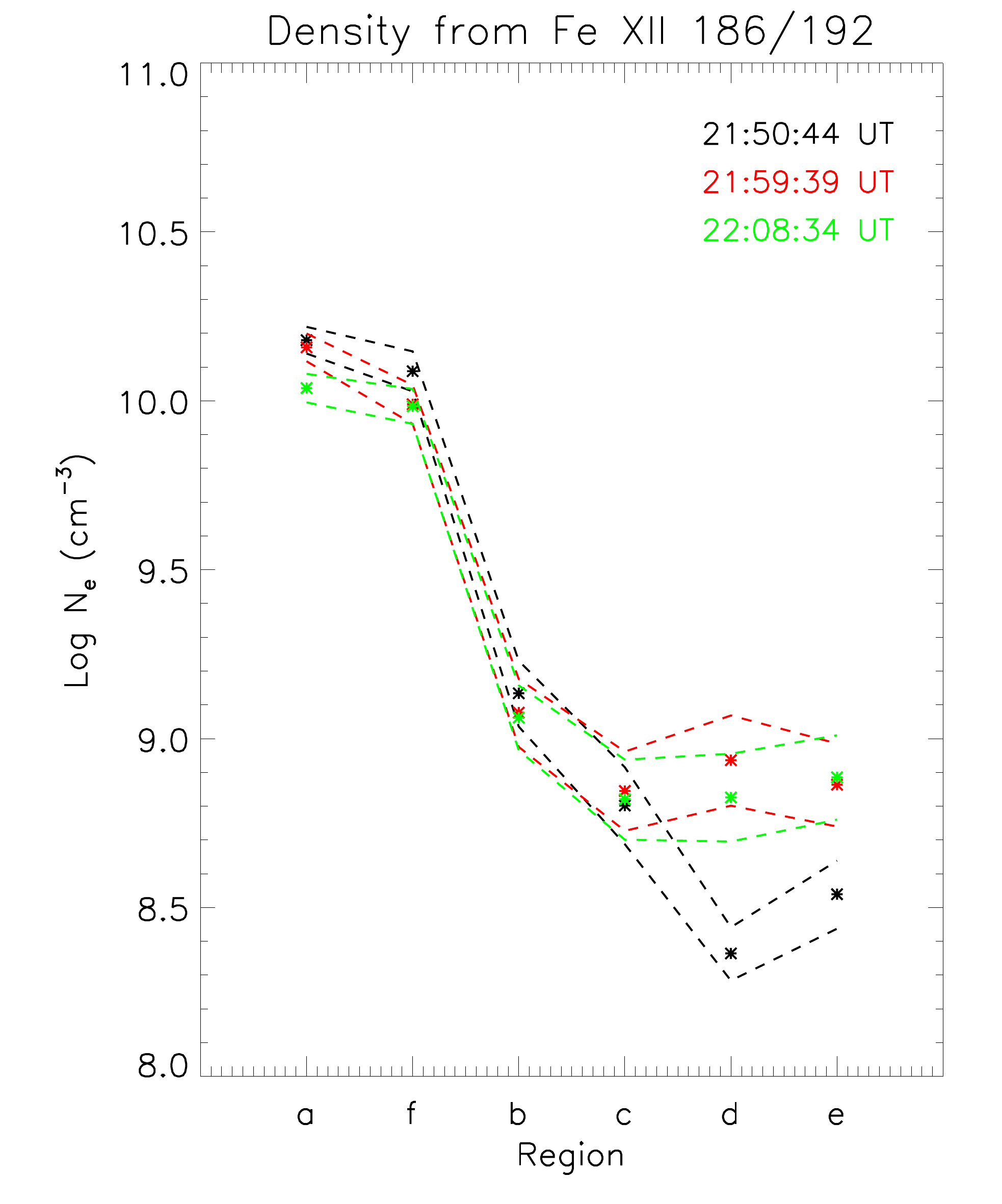}
    \caption{Electron density measurement from the intensity ratio between  \ion{Fe}{12} (186.88\,\AA) and (192.37\,\AA). The average densities are measured from the same regions (A$-$F in Figure~\ref{fig7}). The dashed lines show the upper and lower limits of the density measurements, which are calculated from the 1-$\sigma$ errors on the intensities.}
    \label{fig13}
\end{figure}

We measured the electron density of the flaring loop region using the density sensitive line pairs of \ion{Fe}{12} (186.88\,\AA) and (192.37\,\AA), which has a peak formation temperature of $\log T=6.2$. Figure~\ref{fig13} shows the density variation with time and location. The measurements along the flaring loop locations reveal that the density decreases with height and varies in the range of $\log N_{e}=8.3-10.2$.  The temporal variation of the density in the loop-top (regions D and E, Figure~\ref{fig7}) indicates that the observed density at the loop-top is enhanced in the second raster ($\sim1.3\times10^{9}$\,cm$^{-3}$) compared to the first raster of the flare start timing ($\sim0.3\times10^{9}$\,cm$^{-3}$). The density enhancement measured from the \ion{Fe}{12} spectra in the second raster implies that the hot plasma density at the loop-top is later observed in \ion{Fe}{12} after the hot plasma cools down. The densities of the loop footpoint and leg do not vary significantly. The average loop-top density at 21:59~UT is $\sim2\times10^{9}$\,cm$^{-3}$, which is used for the cooling time calculation in Section~\ref{subsec:heating-cooling}. 

\clearpage

\subsubsection{Line width} \label{subsubsec:nonthermal_vel}

We also checked the variation of the line width in the flaring loop. The observed spectral line width consists of the thermal and non-thermal broadening combined with instrumental broadening. We measured the non-thermal broadening by assuming that the instrumental broadening of {\it Hinode}/EIS is $\sim0.05-0.068$\,\AA\ and calculating the thermal broadening from the peak formation temperature \citep{young_2011, brooks&warren_2016}. The non-thermal broadening in EUV and X-ray emission is usually regarded as a manifestation of unresolved mass motions of the plasma, such as multiple flows, turbulence, or waves \citep{dere&mason_1993, chae_etal1998b, hara_etal2011, kawate&imada_2013, doschek_etal2014}.

\begin{figure}
\epsscale{1}
\plotone{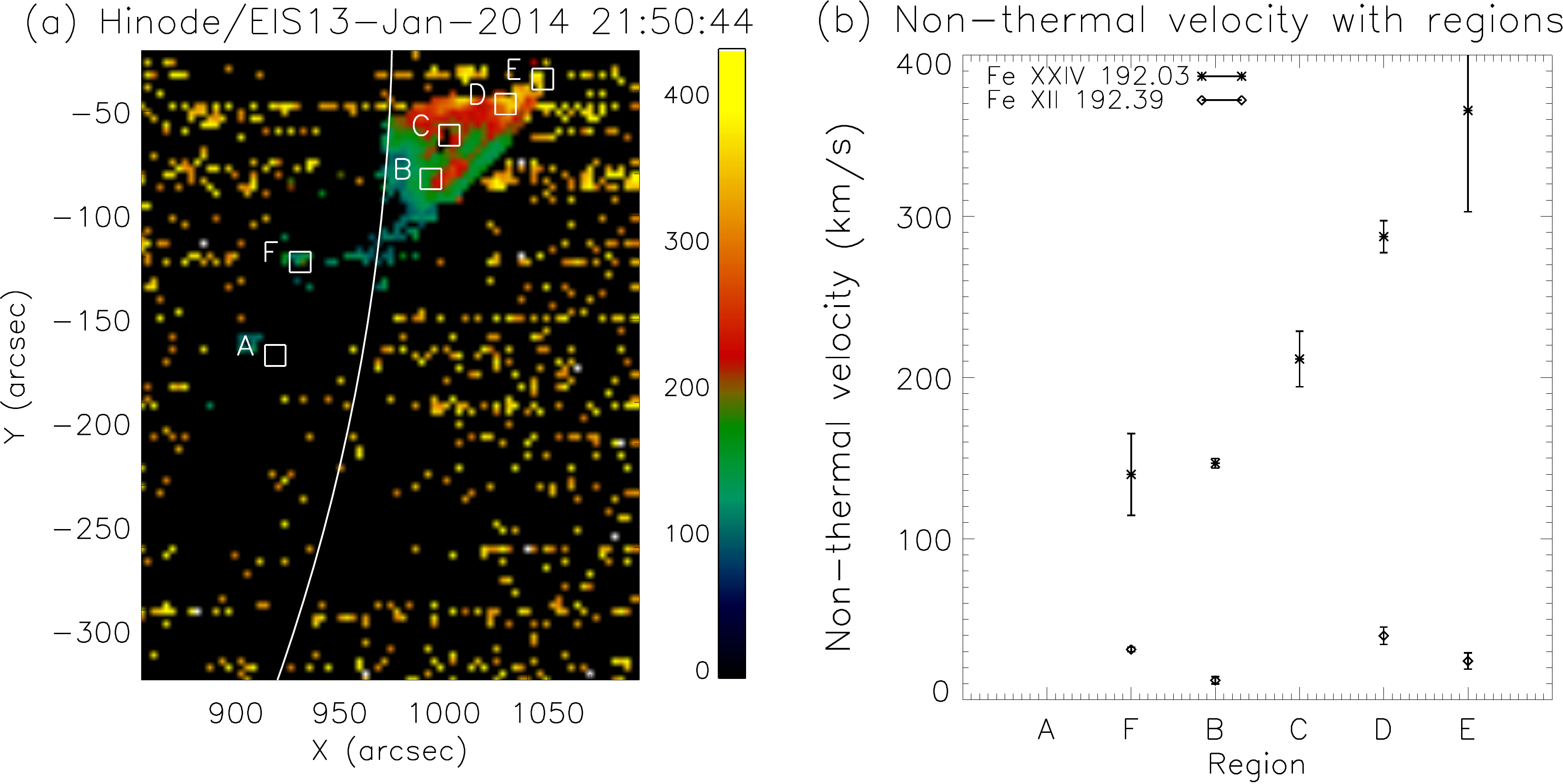} \caption{(a) Non-thermal velocity map at the spectral line of \ion{Fe}{24} 192.03\,\AA\ during the impulsive phase (21:50~UT). The boxes indicate the same location of the boxes in Figures~\ref{fig2} and \ref{fig7}. (b) The non-thermal velocities derived from the \ion{Fe}{24} (asterisk) and \ion{Fe}{12} (diamond) spectral lines for each box locations in panel (a).  \label{fig14}}
\end{figure}

Figure~\ref{fig14} (a) shows the non-thermal velocity map during the flare at the spectral line of \ion{Fe}{24} (192.03\,\AA). Strong non-thermal velocity is only observed in the loop-top region. The spatial and temporal variations of the \ion{Fe}{24} and \ion{Fe}{12} non-thermal velocities are provided in panel (b). For regions A and C (Figure~\ref{fig7}), the observed width from \ion{Fe}{12} is less than the sum of the thermal broadening and instrumental broadening, so the non-thermal velocities cannot be calculated. Also, \ion{Fe}{24} emission does not appear in region A. Figure~\ref{fig14} shows that the non-thermal velocity in \ion{Fe}{24} increases upward along the loop with values ranging from 100$-$350\,km\,s$^{-1}$ and is greatest at the loop-top region. However, the non-thermal velocity from \ion{Fe}{12} ranges from  10$-$40\,km\,s$^{-1}$ and does not vary with height. If magnetic reconnection occurs near the loop-top region, the non-thermal broadening is possibly caused by turbulent motion or waves due to the reconnection outflows and/or accelerated electrons \citep{hamilton&petrosian_1992, liu_etal2008, susino_etal2013}. 

We note that the non-thermal velocity value in our analysis is considerably larger than previous studies \citep{doschek_etal2014, susino_etal2013}. However, there have been several current sheet studies that have reported similarly large values of non-thermal velocity, $\sim150-350$\,km\,s$^{-1}$ \citep{harra_etal2001, cirtain_etal2013, li_y_etal2018}. Moreover, MHD simulations performed by \citet{gordovskyy_etal2016} suggest that the higher temperature plasma may produce larger non-thermal velocities. Specifically, $\sim200-400$\,km\,s$^{-1}$ of non-thermal velocity enhancement is expected near 10$-$20\,MK plasma due to the correlation between the flow velocities, velocity dispersions, and temperatures.  Our highest measured non-thermal velocities extend up to $\sim350$\,km\,s$^{-1}$ (the highest value has a large error bar) and are indeed associated with 10$-$20\,MK plasma. It appears to be caused by the large Doppler velocity of the loop-top plasma. The superposed multi-loop strands, resulting from the curved loop structure configuration, may also contribute to the large width. 

In addition, enhanced intensity from strong flows can also increase the line width. We examined the temporal intensity variation along the loops to determine if this enhanced intensity could be generating misleading line broadening measurements. The results demonstrate that the intensity is not enhanced along the loop during the impulsive phase. Moreover, the intensity enhancement from flows (see Figure~\ref{fig6}) occurs during the gradual phase, which corresponds to a strong decrease in the non-thermal velocity. This anti-correlation implies that the enhancement of non-thermal velocity in the impulsive phases is likely caused directly by strong flows or turbulent motions, rather than a systematic increase due to intensity enhancement.

\clearpage

\subsection{Stereoscopy}\label{subsec:stereoscopy}

If the flaring loop is located along the plane of the sky, the line of sight velocity of the moving plasma would be not significant. However, if the loop is tilted from the plane of the sky, the velocity component of the line of sight would increase with the tilt angle. The observed strongly red-shifted emission in this flaring loop could be interpreted as strong upflows or downflows along the loop depending on the angle between the loop and the plane of the sky. If the loop leg tilts away from the observer, the red-shifted emission would be interpreted as an upflow along the loop. Conversely, if the loop leg tilts toward the observer, the  red-shift would be considered emanating from a downflow. We do not know the loop orientation because it is projected at the limb, and the tilt angle of the loop will determine the direction of flows (i.e., upflow or downflow). Therefore, the loop tilt angle and 3-D direction are critical for removing this degeneracy in the Doppler velocities in order to understand where the reconnection occurs and how the flows originate. 

\begin{figure*}
\epsscale{1.}
\plotone{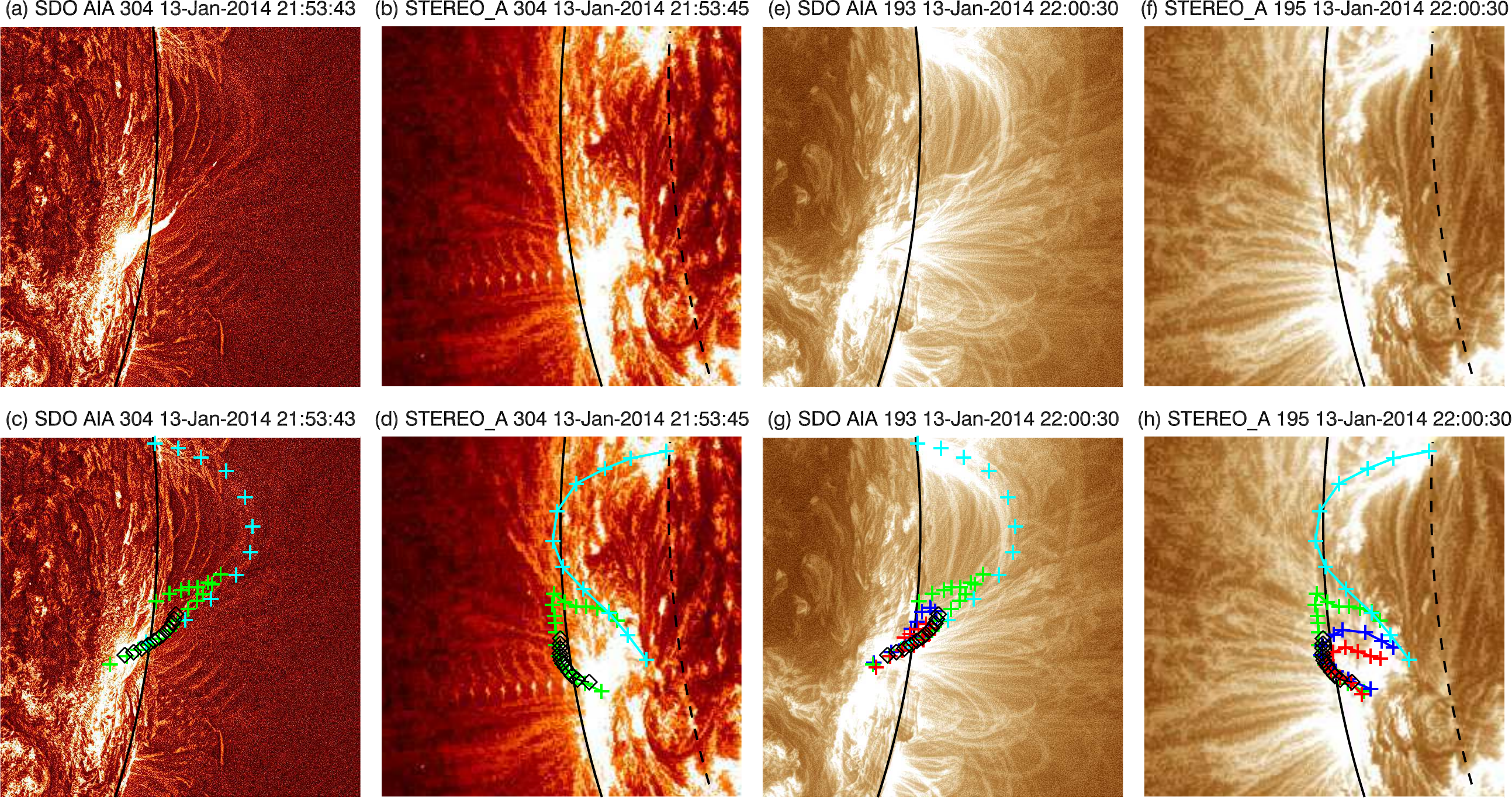} \caption{{\it SDO}/AIA and {\it STEREO}-A/EUVI images with the loop position identified using the {\tt scc\_measure.pro} mapping routine. (a$-$d) AIA and EUVI 304\,\AA\ images at the flare start. The black diamonds trace the cool jet structure from the 304\,\AA\ images. The cyan and green crosses depict the trans-equatorial loop and cusp-shape loop traced from the AIA 193\,\AA\ and EUVI 195\,\AA\ images. (e$-$h) AIA 193\,\AA\ and EUVI 195\,\AA\ images during the flare. The blue and red crosses track the flare arcade and the path of the jet, which is seen in both the AIA 193\,\AA\ and EUVI 195\,\AA\ passbands. The solid black lines indicate the limb location in each of the images. The dashed black lines correspond to the AIA limb location projected onto the EUVI coordinates. \label{fig15}}
\end{figure*}

\begin{figure}
\epsscale{1.}
\plotone{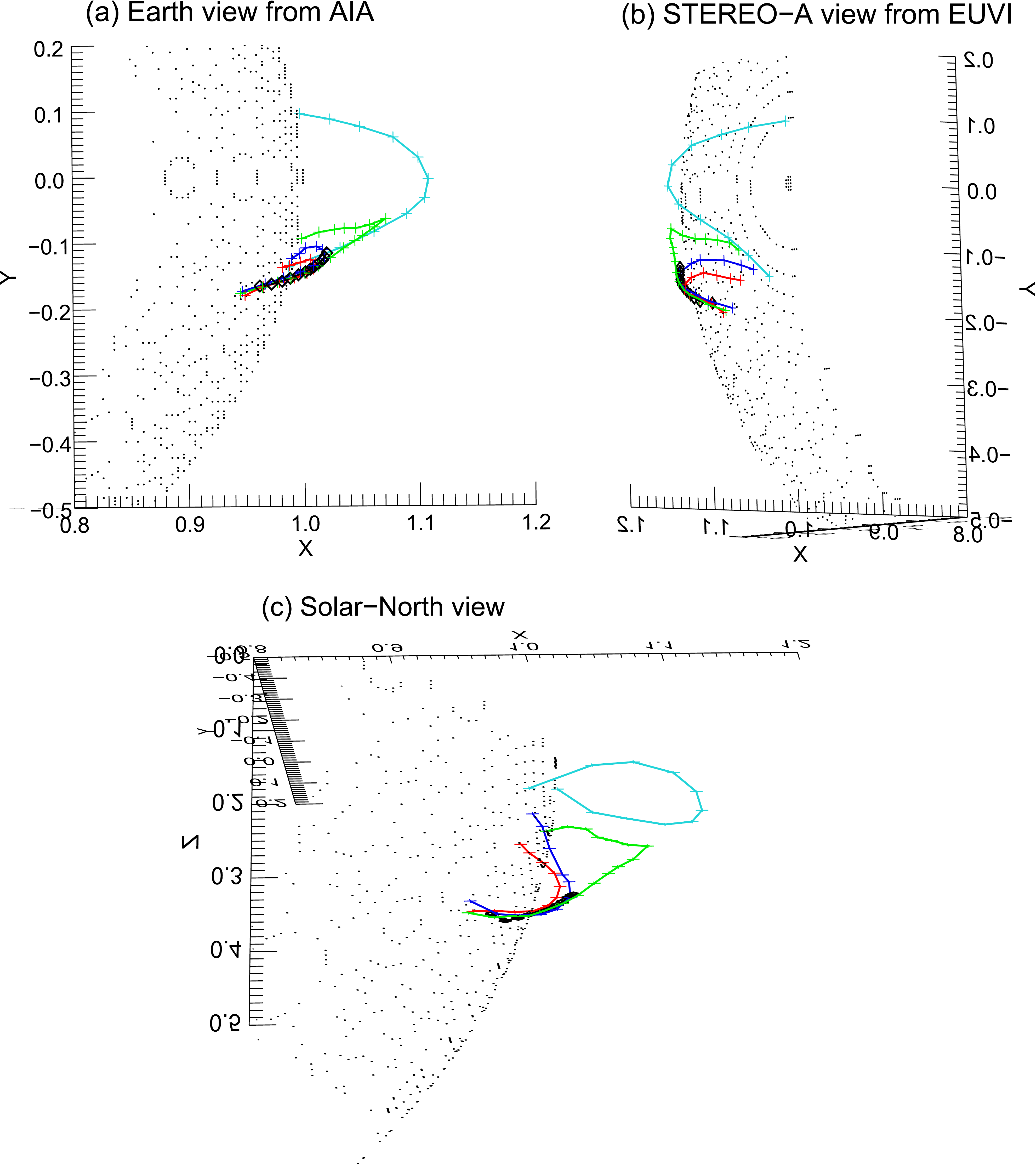} \caption{The 3-D reconstructed loops using the loop traced positions shown in Figure~\ref{fig15}. (a) {\it SDO} view, (b) {\it STEREO}-A view, and (c) solar north view. The colors and symbols of the traced loops are same as Figure~\ref{fig15}.  \label{fig16}}
\end{figure}

Taking advantage of the multiple vantage points between the $\it{STEREO}$/EUVI and $\it{SDO}$/AIA observations, we reconstructed the 3-D path of the flaring loop. Figure~\ref{fig15} shows nearly co-temporal 195 and 304\,\AA\ images from both EUVI and AIA. The EUVI and AIA images are processed via Multi-Scale Gaussian Normalization to clearly show the loop path \citep{morgan_druckmuller_2014}. To trace the loop, we employ the stereoscopic reconstruction technique, {\tt scc\_measure.pro} \citep{thompson_2009} to determine the 3-D coordinates of these points in the heliographic system. Using an equipotential line between two different angle images, points are selected along the jet in the 304\,\AA\ channel (black diamond) as well as along the flaring loops in the 193 and 195\,\AA\ channels (blue and green crosses). The detailed analysis is available in the Appendix. Using the points along the loop, we determined an average tilt angle of 51$^{\circ}$ for the plane of the coronal loop with respect to the plane of the solar equator. The calculated 3-D coordinates are displayed in Figure~\ref{fig16}. The panels (a) and (b) display the constructed loop as seen from Earth and from {\it STEREO}-A, respectively. The reconstruction matches well with the observations. Panel (c) is the view from solar north. We find that the direction of the hot loop from its eastern foot to the cusp region (green crosses with line) is tilted away from us. Thus, from the 3-D construction using {\it STEREO} data, we conclude that the observed fast velocity hot plasma component is due to upflowing material along the loop. 

\clearpage

\subsection{Magnetic field extrapolation} \label{subsec:magnetic field}

\begin{figure}
\epsscale{1.}
\plotone{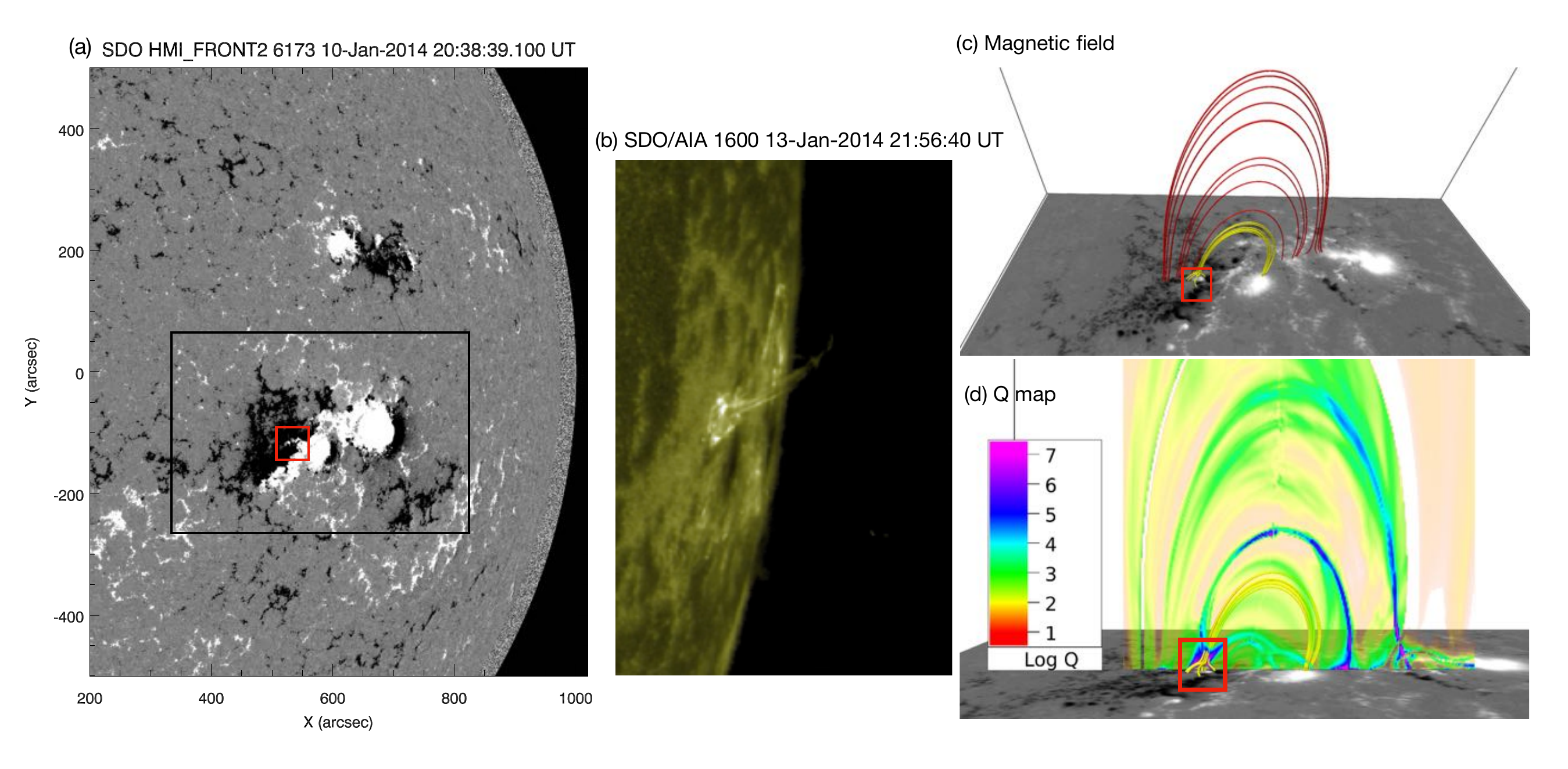} \caption{The result of the magnetic field extrapolation using an {\it{SDO}}/HMI vector magnetogram from three days before the flare. (a) The region of the HMI vector magnetogram used for the input to the magnetic extrapolation, outlined by the black box. The grey-scaled image in the box is the line of sight component of the vector magnetogram. The red box centers on the isolated region of positive-polarity. (b) The {\it{SDO}}/AIA 1600\,\AA\ image shows the sunspot location and flare brigthening at the two ends of the flaring arch. (c) The extrapolated magnetic fields. Red lines indicate the overlying fields while yellow lines correspond to the inner reconnecting fields. (d) The Q-map showing the squashing factor values. The red boxes in panels (c) and (d) mark the jet location. \label{fig17}}
\end{figure}

We extrapolated the coronal magnetic fields using {\it SDO}/HMI vector magnetograms and the NFFF extrapolation code described in \citet{2010JASTP..72..219H}. Since we have used the HMI magnetogram from three days before the limb flare (from when the active region was in view), the magnetic field configuration is typically expected to differ from that at the time of the flare. We note from the observed EUV images, though, that the active region loop configuration does not change much for the next three days. Thus, we assume that the magnetic configuration also remains mostly unchanged. Then, by comparing the HMI magnetogram with {\it SDO}/AIA 1600\,\AA\ images carefully, we identified the location of the jet ejection related to the flare, which is a localized region of positive polarity isolated within the active region's globally dominant negative polarity (the red box in Figure~\ref{fig17} (a)).

The vector field shown in the black rectangle of Figure~\ref{fig17} outlines the cutout used for the extrapolation, with dimensions of $1024\times512$ pixels centered at (766.55, -128.63) arcseconds. To reduce the computational cost, we rescaled the original field and used a $512\times256\times256$ pixel grid volume in the $x$, $y$, and $z$ directions (panel~(c)). The $x$, $y$, and $z$ vectors refer to the local Cartesian coordinate system, which is obtained after the active region has been de-projected and de-rotated to the disk center. The $z$-direction is the line of sight, and the transverse field is in the $x-y$ plane. 

The average deviation between the observed ($\mathbf{B}_t$) and the extrapolated ($\mathbf{b}_t$) transverse field on the photospheric boundary is indicated by the following metric:
\begin{equation}
E_n =\left(\sum_{i=1}^M |\mathbf{B}_{t,i}-\mathbf{b}_{t,i}|\times|\mathbf{B}_{t,i}|\right)/\left(\sum_{i=1}^M |\mathbf{B}_{t,i}|^2\right)
\label{en}
\end{equation}
where $M=N^2$, represents the total number of grids points on the transverse plane.
In order to minimize the contribution from the weaker fields, the grid points are weighted with respect to the strength of the observed transverse field.
Since $E_n$ represents the normalized vector error between $\mathbf{b}_{t}$ and the measured $\mathbf{B}_{t}$, a perfect correlation will give $E_n=0$. In the present case, the error in the weighted transverse field corresponds to $E_n= 0.21$, which is acceptable for the current analysis.

For further investigation of the magnetic field topology, we look at the regions of strong variation in the field line connectivity \citep{1996A&A...308..643D}. Such regions, having a high value of the squashing factor Q, are thought to be the plausible places for magnetic reconnection to occur \citep{1995JGR...10023443P}. Here, the squashing factor is calculated by following \citet{2016ApJ...818..148L}.

The extrapolated magnetic field and the Q map also show that there is an inner loop structure (yellow color in Figure~\ref{fig17}~(c)), and its field line connectivity changes sharply at the location between the inner loop and surrounding field lines (having high Q values). This configuration implies that the jet ejection originates from magnetic reconnection in the null region above the isolated region of positive polarity with the fan-spine magnetic topology. In addition, we observe a remote brightening at the surface in the AIA 1600\,\AA\ images concurrent with the jet and flare brightening (Figure 2). The location of the other loop footpoint from the extrapolated magnetic field corresponds to this remote brightening location, which confirms that the extrapolated field is consistent with the EUV observation. 

\begin{figure}
\epsscale{0.8}
\plotone{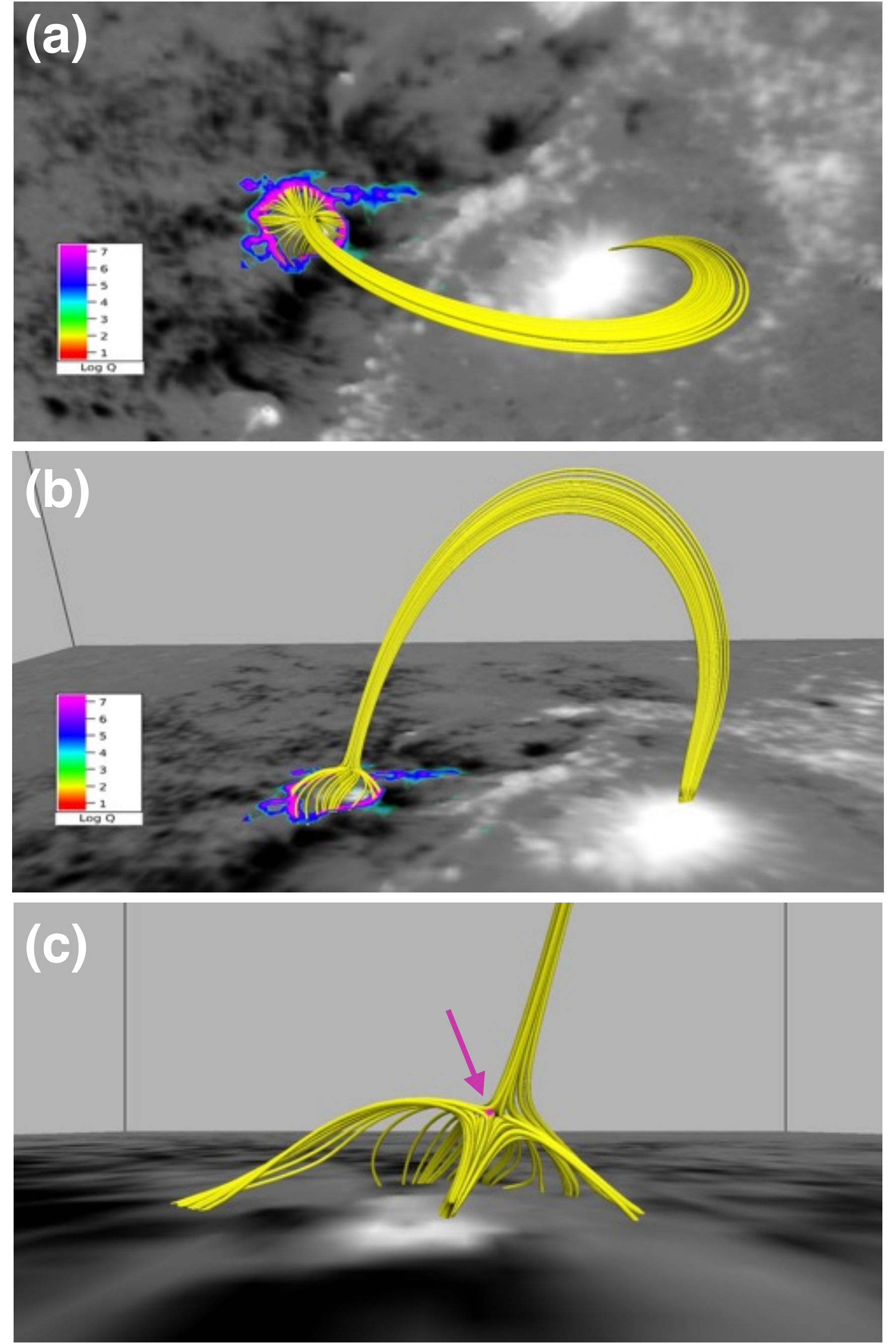} \caption{Enlarged images of the extrapolated field lines and Q map for the inner reconnecting field (yellow field lines from Figure~\ref{fig17}). (a) Top view. (b) Side view. The colored bottom boundary shows contours of high value of log Q. (c) Extracted view of the fan-spine structure on the $x-z$ plane. The purple arrow points to the null point structure having high value of log Q (small colored contour). \label{fig18}}
\end{figure}

Figure~\ref{fig18} shows zoomed-in images of the extrapolated field lines and Q map for the jet reconnection region (panel (a): top view; panel (b): side view). The field lines and the circular bottom boundary of contours of high log Q value in the jet base suggest that this region has a fan-spine structure. Panel (c) is an extracted view of the jet-base on the $x-z$ plane showing where the highest Q value is observed. The small, narrow (purple) region having the highest Q value is the most plausible reconnecting null point between the erupting field and the magnetic fan arch.

\begin{figure}
\epsscale{1.}
\plotone{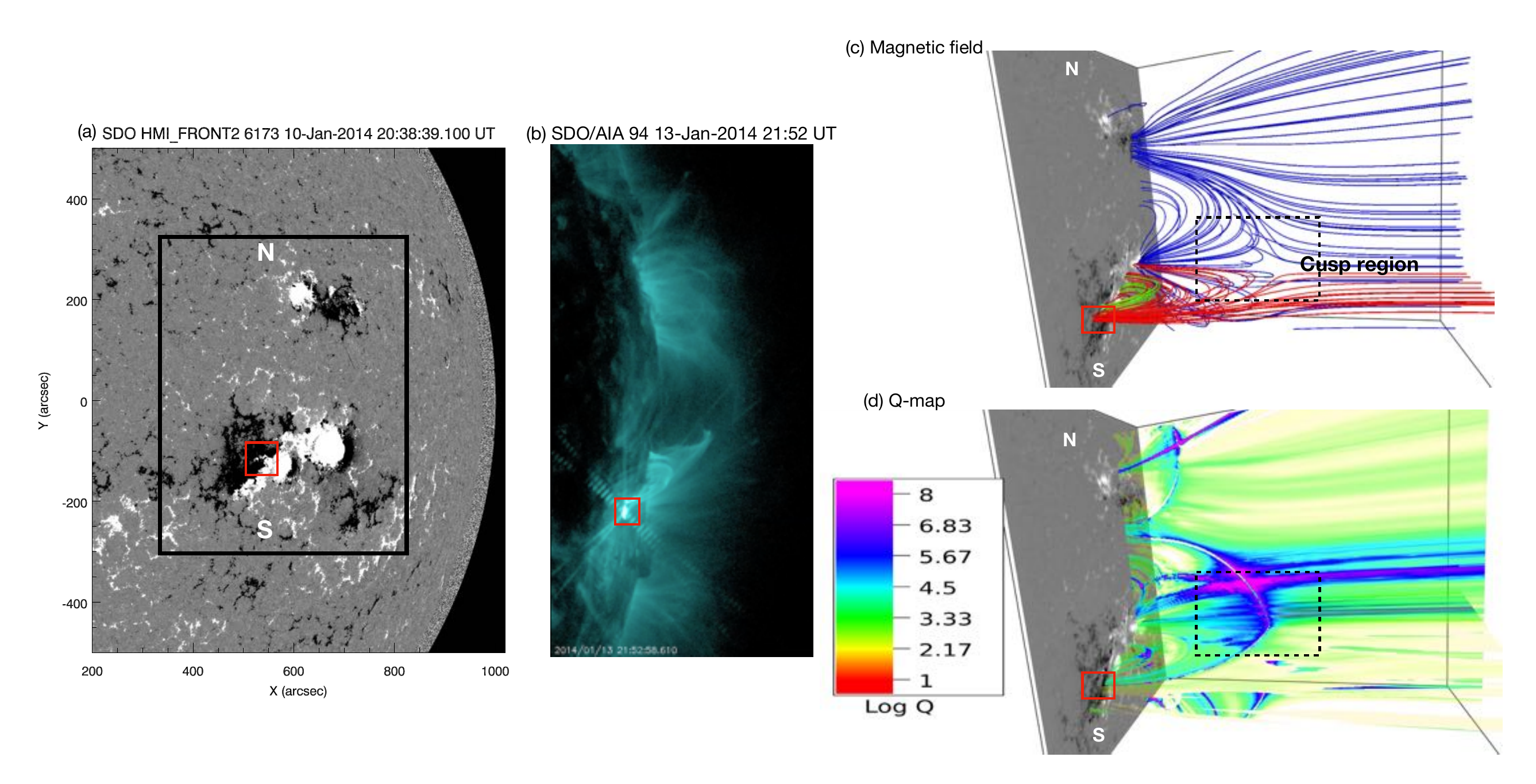} \caption{The result of the NFFF extrapolation using the {\it SDO}/HMI vector magnetogram with a large FOV including a trans-equatorial loop. (a) The region of the HMI vector magnetogram used for the input to the magnetic field extrapolation, outlined by the black box. The red box centers on the isolated region of positive-polarity at the jet location. (b) The {\it{SDO}}/AIA 94\,\AA\ image shows the hot flaring loop and trans-equatorial loop. (c) The extrapolated magnetic fields. There are two adjacent loop systems. Green and red lines correspond to the inner and overlying loop field lines in Figure~\ref{fig17}. The blue lines are the trans-equatorial loop field lines. (d) The Q-map showing the squashing factor values. The red and black dashed boxes in panels~(c) and (d) indicate the jet location and the cusp loop top region, respectively. \label{fig19}}
\end{figure}

To understand the connectivity with the trans-equatorial loop that appears to be connected to the loop-top region, we take a magnetogram cutout from January 11 at 23:00~UT with dimensions of $1000\times1200$ pixels centered at (774.8, 23.25) arcseconds (Figure~\ref{fig19} (a)). This extraction was rescaled to a $256\times300\times300$ pixel grid volume in the $x$, $y$, and $z$ directions. In this case, we get $E_n = 0.3$.

Figure~\ref{fig19} shows the extrapolated magnetic field (lower resolution) and the Q map for the flaring active region and the active region at the north end of the trans-equatorial loop. From the AIA 94\,\AA\ image, we can see the cusp shape of the flaring loop top region and the trans-equatorial loop, which appears to be connected with the two loop systems. Our extrapolated magnetic field shows that those two loop bundles are close to each other, and that the top of the fan loop could be seen as the observed cusp shape at the limb. Furthermore, the cusp-shape region also has high Q value, implying that this location is also a plausible place for  magnetic reconnection.

\clearpage

\section{Discussion} \label{sec:discussion}

The limb flare we have explored shows the hot cusp structure in the corona, which is the typical feature in the standard eruptive flare model. We searched for a CME associated with this flare eruption to determine whether it was eruptive or confined. While a partial halo CME occurred near the time of this flare, there was also an eruptive flare located on the backside disk center that was observed by {\it STEREO}-A/EUVI. Considering the large angular width and direction of the eruption, the CME was most likely caused by the backside flare event. Therefore, the limb flare eruption that we have investigated is confined by the overlying magnetic-fan-shaped loop as seen in Figures~\ref{fig17} and \ref{fig19}. Both eruptive and confined flare characteristics of this limb flare have also been reported by \citet{Hernandez_Perez_2019}. Considering the flare plasma properties that we have presented combined with an extrapolated magnetic field configuration, we discuss how the flare is formed and how it could produce the hot X-ray cusp with strong upflow.

\subsection{Upflows in the hot cusp-shape loop-top: Reconnection outflows? or evaporation flows?} \label{subsec:plasmaflows}

Using multi-wavelength observations of the limb flare from two unique vantage points, we find a hot cusp-shaped structure at the fan loop/arch apex exhibiting strong upflows during the impulsive phase. Strong plasma flow during flares is typical and is often interpreted as a response of the plasma to energy release and heating. One possible flow source is reconnection outflow (i.e.,  bidirectional flows originating from the reconnection point traveling near the Alfv\'en speed) \citep{wang_etal2007, tian_etal2014}.  Another possibility is evaporation flow, which is heated plasma ejected from the chromosphere due to energy injection from accelerated particles \citep{milligan&dennis_2009, polito_etal2016b, lee_etal2017}. 

To understand whether the flow to the loop-top region is reconnection outflow or evaporation flow, we must consider the temperature dependence of the flow velocity. If the flows were due to evaporation, which is the result of a pressure difference due to heating, their speeds would be dependent on temperature and would be comparable to the sound speed. The measured temperature in Figure~\ref{fig12} suggests that evaporation flows would be $\sim370-550$\,km\,s$^{-1}$ for temperatures of 10$-$22\,MK, and the velocity would increase with temperature. The speeds calculated from the {\it Hinode}/EIS Doppler velocity measurements combined with the derived loop tilt angle from {\it STEREO}-A observations show that the de-projected observed velocities are consistent with the the evaporation flow scenario.

For the upflows to be reconnection outflows, the reconnection region needs to be located below the cusp region. The temperature measurements along the loop indicate that the hottest region is the curved loop structure (region C in Figure~\ref{fig2}), just below the upflowing plasma. The spectral line profile at the curved loop shows that this hot plasma has blue- and red-shifted bidirectional flows. Assuming a coronal magnetic field strength of 30\,G \citep{nakariakov&ofman_2001, Van_Doorsselaere_etal2008, jess_etal2016}, the Alfv\'en speed would be about $\sim1000-3000$\,km\,s$^{-1}$ for the measured density of $\sim10^9-10^{10}$\,cm$^{-3}$. The observed velocity is slower than the expected Alfv\'en speed. 

\subsection{Cooling process after the impulsive heating in the flare}\label{subsec:heating-cooling}

Our targeted flare for this study is impulsive with a duration of only $\sim$5~minutes in the {\it GOES} SXR light curve. This impulsive heating is in contrast to long duration events in which the decay phase has continued heating for many hours \citep{reeves&warren_2002, warren_2006, savage_etal2012} or a second phase of robust heating \citep{qiu&longcope_2016, zhu_etal2018}. For this impulsive flare case, we expect for the theoretical models of loop cooling timescales to be consistent with the observed cooling times. We test this hypothesis by comparing the theoretical cooling times with the observed temporal evolution.

\begin{figure}
    \epsscale{0.9}
    \plotone{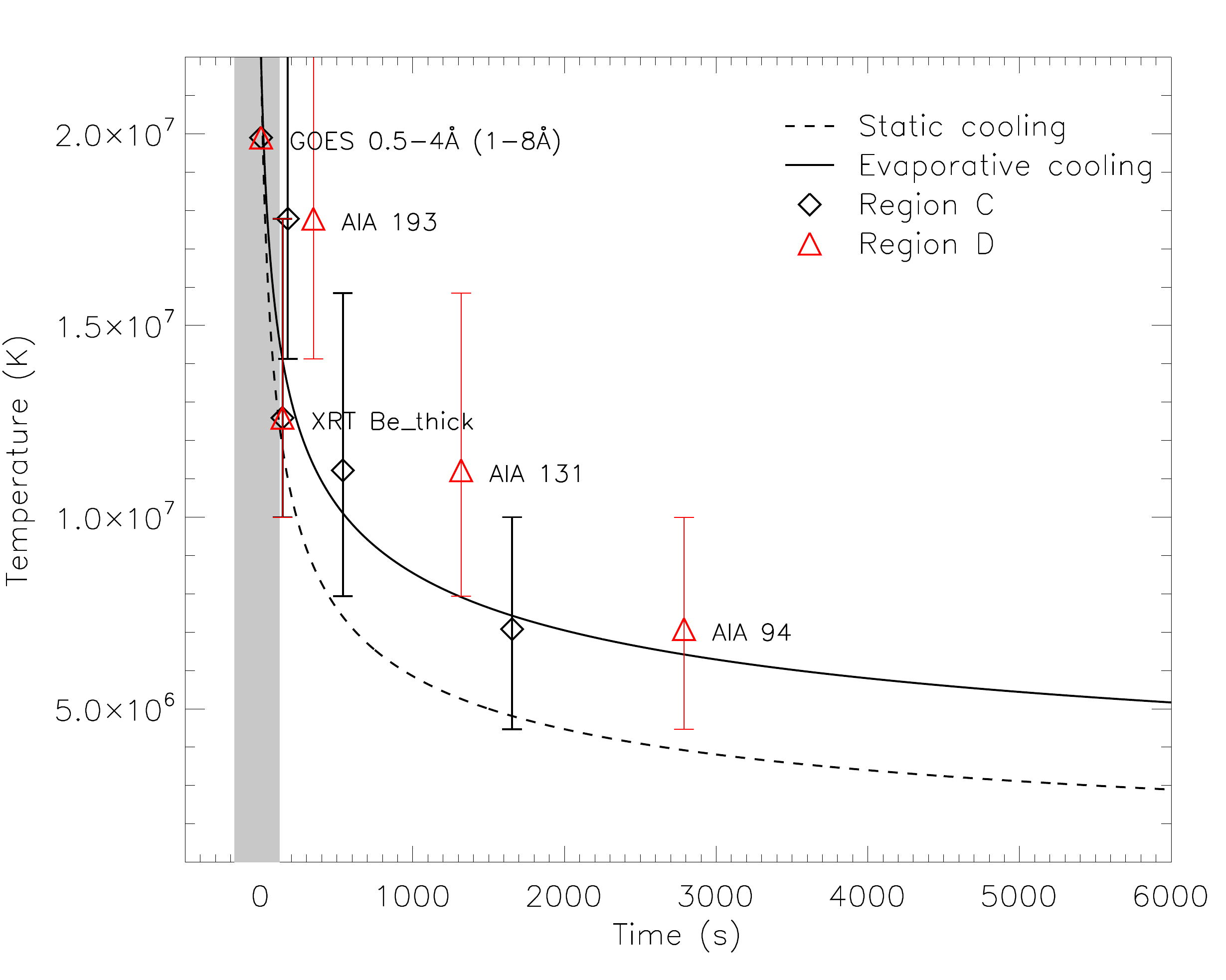}
    \caption{Temporal evolution of temperature estimated from the theoretical conductive cooling models.Dashed and solid lines correspond to the static and evaporative cooling case, respectively. Diamond and triangle symbols are the observed peak time delays from the different instrument and wavelength channels using their peak response temperatures for regions C and D in Figure~\ref{fig2} (see Table~\ref{tbl-2}). The error bars indicate the half width of the temperature peak of the response functions for each channel. The grey area indicates the jet flare duration from the {\it GOES} SXR light curve in Figure~\ref{fig1}. \label{fig20}}
\end{figure}

We calculated the conductive ($\tau_{C}$) and radiative ($\tau_{R}$) cooling timescales for the flaring loop \citep{cargill_etal1995, bradshaw&cargill_2010}, using: 
\begin{equation}
\tau_{C}=\frac{2 n k_{b} L^{2}}{(\gamma -1) \kappa_{0} T^{5/2}} ~\mathrm{and} ~ \tau_{R}=\frac{2 k_{b} T^{1-\alpha}}{(\gamma -1) \chi n}
\end{equation}
where $k_{b}$ is Boltzmann constant, $\kappa_{0}$ is the coefficient of thermal conduction, and $\gamma=\frac{5}{3}$. The optically thin radiative loss function of $R=n^{2} \chi T^{\alpha}$ is referred from equation (3) in \citet{klimchuk_etal2008}. Using the half-length of the loop from the magnetic field extrapolation ($L=0.9\times10^5$\,km) plus the measured density ($n=2\times10^{9}$\,cm$^{-3}$) and temperature ($T\sim22$\,MK at the loop top) from the EIS observations, the calculated conductive and radiative cooling timescales for the flare plasma at the loop-top are approximately 38\,s and 180000\,s, respectively. The conductive cooling is dominant in this flare loop due to its high temperature and low density. Then, we fitted the temporal evolution of the temperature using the conductive cooling model for static and evaporative cooling (equations (3a) and (3b) from \citet{cargill_etal1995}). These fits are provided in Figure~\ref{fig20}.  

The observed temporal evolution of temperature in the flare plasma is measured from the {\it GOES}, XRT, and AIA light curves (Figures~\ref{fig1} and \ref{fig5}). These emission profiles reveal that the intensity peak has a delay from hotter temperature channels to cooler ones. Table~\ref{tbl-2} gives the peak times of the intensity for the near loop-top regions (C and D in Figure~\ref{fig5}) per channel. The time delays relative to the {\it GOES} SXR emission and the peak response temperature of the high temperature channels are also provided in the table. We overlaid the observed evolution of cooling plasma (triangle and diamond symbols) in Figure~\ref{fig20}. The peak temperature of each channel is determined from its dominant spectral lines \citep{odwyer_etal2010, odwyer_etal2014}. We also plotted the temperature range using the half width of the peak temperature response function as an error bar since the channels are composed of more than one line and the response function is broad.

Overall, taking into account the broad temperature response function, the observed temperature evolution is consistent with the conductive cooling curve for the evaporation (solid line) rather than the static (dashed line) case. The observed evolution of the temperature in the flaring loop arcade (region C) agrees well with the model. This result implies that there was insignificant additional heating of the flaring fan loop after the impulsive energy input from the jet. Moreover, it indicates that this flaring loop cooled down by conduction with evaporation.

However, the cusp (region D) experiences a time delay in the cooling curve of $\sim$15\,min for the emission from the AIA 94 and 131\,\AA\ channels (7$-$11\,MK). There could be several reasons for the observed longer cooling time for the flare cusp. First, continued reconnection of substrands of the fan loop could extend the cooling time \citep{reeves&warren_2002, warren_2006}. But in this scenario, we also expect a longer cooling time in region C since it is also part of the flaring fan loop. Second, there might be a prolonged heating from secondary reconnection at the cusp region after the initial impulsive phase \citep{qiu&longcope_2016, zhu_etal2018}. \citet{Hernandez_Perez_2019} studied the same flare and proposed that there would be slow local reconnection near the kinked loop apex due to the increase of the thermal pressure by continuous plasma upflows. The grey area in Figure~\ref{fig20} shows the {\it GOES} flare duration when the jet ejection appeared. The impulsive heating from the jet contributes to the {\it GOES} SXR spike and the creation of the hot plasma ($>10$\,MK). The additional heating process would be responsible for the declining phase of the flare SXR flux seen by {\it GOES} and the delayed intensity peaks of the AIA 131 and 94\,\AA\ light curves from the cusp region.

\clearpage

\subsection{Magnetic-fan flaring arch: Relation between the magnetic topology, cool jet ejection, X-ray hot outflow, and remote footpoint brightening}

The extrapolated magnetic fields and their connectivity show that there are two possible places where magnetic reconnection might take place. One is near the cool jet ejection site where footpoint brightening is observed (Figure~\ref{fig17}). The other is the cusp region above the loop-top where the two loop systems are in contact (Figure~\ref{fig19}).  Reconnection at either of these locations could plausibly convert sufficient amounts of magnetic energy into thermal and kinetic energies to produce this flare. 

If we consider that the higher coronal magnetic reconnection at the cusp region produces the flare, as prescribed by the standard eruptive flare model, the observed non-thermal velocity and temperature measurements are consistent with the model expectation \citep{doschek_etal2014}. Also, the hot and upflowing plasma in the cusp region can be interpreted as the newly formed current sheet or evaporation flows by injection of energy from the reconnection. However, if there is such continuous reconnection above the loop-top, we expect HXR emission at the loop-top region. No HXR emission was observed at the loop-top when the flare starts, and the observed cooling time implies that the flaring loop was impulsively heated with no evidence for continued heating. We note that the loop cusp experiences a longer cooling time for 7$-$11\,MK plasma in the decay phase of the flare, but this process requires an additional heating mechanism separate from that of the impulsive phase, perhaps by secondary reconnection at the cusp. 

Another possibility supported by all of these observations is that the external reconnection of the blowout jet (i.e., external (interchange) reconnection taking place at the jet eruption region) energized this flare. The cool jet ejection in the AIA 304 and 1600\,\AA\ images and the enhanced {\it{RHESSI}} HXR emission at the flaring loop footpoint at the base of the jet support this scenario. If we consider that the external reconnection produced the jet spire and heated the flare plasma, the HXR emission should only be observed at the footpoint, not the loop-top region, which is consistent with our observation. Following from this hypothesis, the observed temperature and velocity structure above the loop-top region would be interpreted as the evaporation flow and outflows driven by the lower atmospheric reconnection. The SXR emission enhancement at the flare loop-top can also be explained as the increased density from the evaporated  plasma that results from the accelerated particles driven from the chromosphere by the jet reconnection.

A comprehensive set of observed features consists of the following: 1) primary footpoint brightening with HXR emission, 2) lack of HXR emission at the loop-top, 3) upflowing hot plasma along the loop-top, 4) enhancement of the loop-top SXR emission after the footpoint brightening, 5) short duration of the event in the {\it GOES} SXR light curve, and 6) the magnetic topology characterized by an overlying magnetic-fan loop system. We infer from this feature set that this flare is an example of a flaring arch \citep{martin&svestka_1988, svestka_etal1989, fontenla_etal1991, hanaoka1997, moore_etal1999, moore&sterling_2007} that is energized by external reconnection associated with an underlying blowout jet eruption. Flaring arches have been reported as a particular component of some flares \citet{martin&svestka_1988, svestka_etal1989, fontenla_etal1991}. This class of events has chromospheric elements at one foot of a coronal flare loop with traversing X-ray and H$\alpha$ emission. The hot plasma in flare arches is confined in the body of the magnetic-fan arch structure, and the accelerated electrons and hot X-ray emission propagate through the arch from its explosive foot. 

\begin{figure}
\epsscale{0.8}
\plotone{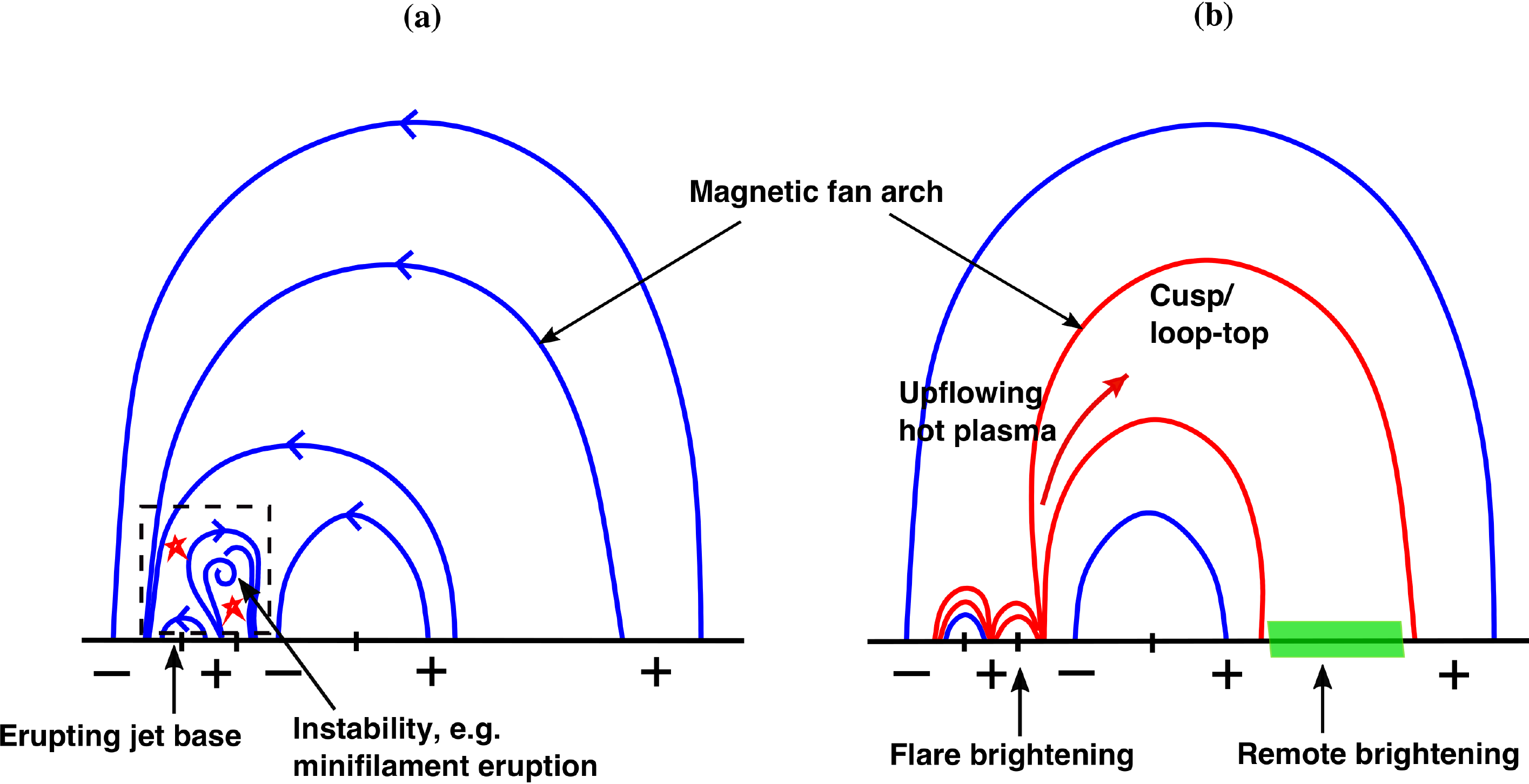} \caption{Schematic cartoon of the blowout-jet and the magnetic-fan flare arch model for this flare. The star symbols mark the magnetic reconnection sites. The blue and red lines correspond to the magnetic field configuration. Panel (a) depicts the topology and reconnection of the magnetic field in the eruption that makes the jet and heats the magnetic-fan flaring arch by driving external reconnection of the erupting field with field in the leg of the fan. Panel (b) depicts how the external reconnection expels accelerated electrons and heated plasma (red arrow) along the reconnected field of the magnetic-fan arch (red field lines). The chromospheric evaporation component is not shown. The green colored area corresponds to the remote brightening (the flare ribbon in AIA 1600\, \AA\ channel) at the other foot of the flaring loop/arch.  \label{fig21}}
\end{figure}

Figure~\ref{fig21} provides a schematic cartoon of the blowout jet and the magnetic-fan arch model for this flare. Panel (a) depicts the onset of the external reconnection between the erupting jet-base and extended magnetic-fan arch. The strong brightening at the fan-spine loop footpoint (see Figure~\ref{fig2} (a$-$f) and its animation, and Figure~\ref{fig18}), results from that external reconnection. The blowout jet is observed as a cool plasma ejection in the AIA 304\,\AA\ images. Even though we do not see the sites of the internal and external reconnection directly due to image saturation, we propose that the jet-productive field eruption is a mini-filament eruption. \citet{sterling_etal2015, sterling_etal2016, sterling_etal2017} have proposed mini-filaments as the typical instigating component of blowout jet eruptions. The cool jet plasma (0.05\,MK) and the chromospheric brightening at the base of the  erupting field in the AIA 1600\,\AA\ images (refer to Figure 2 (b$-$d) of \citet{Hernandez_Perez_2019} and its animation) just before the jet eruption are manifestations similar to those of jet-productive mini-filament eruptions in active regions \citep{sterling_etal2017}. Panel (b) depicts the magnetic fields after the external reconnection (observed as the low-lying flare arcade and extended flaring arch). The arrow represents the heated plasma upflows and accelerated electrons that fill the magnetic arch. The cusp is filled with hot plasma observed at \ion{Fe}{24}.

This scenario explains the impulsive behavior of this flare and lack of HXR emission at the loop top region. Previously, \citet{moore&sterling_2007} suggested a related magnetic-arch-blowout scenario for a narrow streamer puff CME. Our observation would be the specific case in which the magnetic arch field lines are strong enough that the eruption remains confined.

We note that there is a remote brightening at the other foot of the flaring loop in the AIA 1600\,\AA\ channel, seen as a remote flare ribbon (Figure 2's animation and Figure~\ref{fig17} (b)). One of the observational characteristics of the flaring arch is a secondary remote brightening due to the propagation of the accelerated electrons and hot plasma along the arch. We estimated the propagation speed of the energy injection from the jet footpoint brightening to the remote brightening. The time difference between the two brightenings is measured from the AIA 1600 and 304\,\AA\  intensity images with cadences of 24\,s and 12\,s, respectively. Loop distance between the reconnection brightening and remote brightening is calculated using the extrapolated magnetic field and the distance between the two footpoints from the AIA 1600\,\AA\ image. The measured time difference is 12~seconds, which is the AIA temporal resolution limit. The distance along the extrapolated magnetic loop structure is $\sim1.8\times10^{5}$\,km. The resulting estimated propagation speed is $\sim1.5\times10^{4}$\,km\,s$^{-1}$, which is $\sim$15 times the Alfv\'en velocity. Similar observational features were reported by \citet{tang&moore_1982}. They showed large flares that had remote H$\alpha$ bright patches with type III reverse slope bursts. They interpreted the H$\alpha$ patches as chromospheric heating signatures caused by the accelerated non-thermal electrons propagating from the flare-reconnection end of the closed loop. Their estimated traveling velocity was $\sim10^{5}$\,km\,s$^{-1}$, much faster than the Alfv\'en speed. The observed speed in our flaring arch also implies that accelerated electrons from the jet reconnection region impacted and heated the other footpoint of the magnetic fan.

\section{Summary \& conclusion} \label{sec:summary}

We have comprehensively investigated an M1.3 flare at the limb by combining multi-wavelength imaging and spectroscopic observations. We focused on three things from the observed results: (1) What is the cause of the hot plasma flow motions, such as evaporation or reconnection; (2) Are the observed temperature and density variations consistent with a flare heating model?; and (3) Considering the observed spectroscopic results with the extrapolated magnetic field configuration and coronal loop morphology, how is the flare formed and the hot X-ray flare arch heated? 

First, the observed temperature-dependent upflow and HXR observations imply that the fast upflows are due to evaporation and outflow from the reconnection in the jet end of the flaring arch. Second, the temporal evolution of the multi-thermal plasma and the density variation are consistent with the evaporative cooling model, and subsequently, the cooling process after the brief impulsive phase of the flare is consistent with nearly all of the heating occurring in the impulsive phase. Then, combining the configuration and global structure of the flaring loop  with observed plasma dynamics and properties, we conclude that the flare and hot loop-top source could be formed by the external reconnection of the jet-productive erupting field \citep{sterling_etal2015, sterling_etal2016, sterling_etal2017}. Finally, the accelerated electrons and the evaporated hot plasma produce the remote flare brightening and hot flaring arch (cusp structure) via upflows. 

Still, reconnection at the cusp region remains a possibility, evidenced by the strong non-thermal velocity at the edge of the cusp and the high squashing factor. However, the delayed appearance of the SXR source at the loop-top region and the impulsive nature of the {\it GOES} flare light curves are inconsistent with continuous reconnection occurring at a current sheet formed at the cusp region. Also, we observed pre-existing hot loop/arch structure before the flare onset (Figure~\ref{fig4} (a$-$e)), which confirms that most of the overlying magnetic-fan arch structure was not formed during the flare by current sheet formation processes higher in the corona. 

Therefore, we propose that the flare was energized by external reconnection associated with the jet eruption at the base of the magnetic fan (see Figure~\ref{fig21} for a schematic representation). Panel (a) depicts the primary (jet-inducing) and secondary (external) reconnection sites, and panel (b) depicts the hot upflows (red arrow) along the magnetic-fan flaring arch (red field lines). The arch system is observed as a hot, cusp-shaped loop due to the projection on the limb. The propagation of the accelerated electrons and hot plasma along the arch produces the secondary remote brightening at the other foot of the flaring loop/arch (green colored region).

Our study presents a well-observed example of a flaring arch including quantitative spectroscopic observations at the limb. The suite of observations indicate that the hot plasma in the flaring arch results from low atmospheric reconnection accompanying a jet ejection, instead of from reconnection above the arch as expected in the standard flare model. That is, the external reconnection below the arch induced by the jet eruption likely energized this M-class flare, not higher coronal reconnection above the loop/arch apex. In addition, the comparison between the cooling models \citep{cargill_etal1995} and our observations yields important implications for flare models. For example, the cooling model that we used does not take into account turbulent suppression of conduction \citep{Bian_2016, Bradshaw_2019}, yet the model results are consistent with the observed cooling. This concordance implies that turbulent suppression is relatively weak, at least for this flare. Moreover, in agreement with \citet{Hernandez_Perez_2019}, the comparison of the cooling time in a different loop region shows that there can be a second, more prolonged heating period during a flare, as the longer cooling time in the loop/arch apex is possibly explained by slow reconnection occurring in the cusp region after the initial impulsive phase.

\acknowledgments
We thank the anonymous referee for helpful comments. Data are courtesy of the science teams of {\it Hinode}, {\it RHESSI}, and {\it SDO}. {\it Hinode} is a Japanese mission developed and launched by ISAS/JAXA, with NAOJ as domestic partner and NASA and STFC (UK) as international partners. It is operated by these agencies in cooperation with ESA and the NSC (Norway). The {\it RHESSI} satellite is a NASA SMEX mission. HMI and AIA are instruments on board {\it SDO}, a mission for NASA's Living With a Star program. This work was supported by the {\it Hinode} Project Office at NASA/MSFC and JSPS KAKENHI Grant Numbers JP25220703 (PI: S. Tsuneta). A.P. acknowledges partial support of NASA grant 80NSSC17K0016 and NSF award AGS-1650854. NKP acknowledges current support from NASA’s SDO/AIA (NNG04EA00C).

\clearpage

\appendix
If $(r, \theta, \phi)$ denote the distance, latitude, and longitude of a point in the heliographic system, then the corresponding Cartesian coordinates, $(x, y, z)$, are obtained using the following conversion formulae. We adopt the Cartesian system wherein $x$ is along solar west, $y$ is along solar north,and $z$ is pointing towards the Earth.
\begin{align}
x =& \;r\,\cos{\theta}\,\sin{\phi} \nonumber \\
y =& \;r\,\sin{\theta} \nonumber \\
z =& \;r\,\cos{\theta}\,\cos{\phi} \label{E:cartXYZ}
\end{align}

Assuming that the loop lies in a single plane, the tilt angle of the loop would be the angle between this plane and the solar equatorial plane. Since we need any three (non-linear) points to determine a plane, in order to determine the tilt angle we select all possible combinations of three points from the selected available points along the loop.

If $P\,(x_p, y_p, z_p)$, $Q\,(x_q, y_q, z_q)$ and $R\,(x_r, y_r, z_r)$ are the three points along the loop, then the normal to the loop plane, denoted by $\eta_1$ (= $\overrightarrow{PQ} \times \overrightarrow{PR}$), is given by,
\begin{align}
\eta_1 \,=\; &((y_q - y_p)(z_r - z_p) - (z_q - z_p)(y_r - y_p))\,\hat{e_x} \;+ \nonumber \\
	     &((z_q - z_p)(x_r - x_p) - (x_q - x_p)(z_r - z_p))\,\hat{e_y} \;+ \nonumber \\
	     &((x_q - x_p)(y_r - y_p) - (y_q - y_p)(x_r - x_p))\,\hat{e_z} \label{E:normal}
\end{align}
where, $\hat{e_x}$, $\hat{e_y}$ and $\hat{e_z}$ are the unit vectors along $x$, $y$ and $z$ axes in the Cartesian system. Similarly, the normal to the solar equatorial plane, denoted by $\eta_2$, is given by $\eta_2 = \hat{e_y}$.

Given the normal vectors of two planes, the angle between them, denoted by $\xi$, is:
\begin{equation}
\xi \,=\; \cos^{-1}\bigg(\frac{|\;A_1\,A_2 + B_1\,B_2 + C_1\,C_2\;|}
			{\sqrt{A_1^2 + B_1^2 + C_1^2}\;\sqrt{A_2^2 + B_2^2 + C_2^2}}\bigg) \label{E:loopangle}
\end{equation}
where, $(A_1, B_1, C_1)$ are respectively the $(x, y, z)$ components of $\eta_1$, and $(A_2, B_2, C_2)$ are those for $\eta_2$. In the special case of solar equator, $A_2 = C_2 = 0$ and $B_2 = 1$.

We selected the flaring loop/arch and the jet positions from the two different view points of AIA and EUVI-A images. Then, we calculated an average tilt angle as 51$^{\circ}$ of the coronal loop with respect to the solar equator from the possible combinations among the positions.

\bibliographystyle{aasjournal}
\bibliography{apj}

\end{document}